\documentclass{aa}
\usepackage[varg]{txfonts}
\usepackage{graphicx}
\usepackage{paralist,xcolor}
\usepackage{natbib}
\bibpunct{(}{)}{;}{a}{}{,} 

\usepackage{graphicx}   
\usepackage{amsmath}    
\usepackage{amssymb}    
\usepackage{pdflscape}
\usepackage{hyperref} 
\hypersetup{colorlinks=true,linkcolor=blue,citecolor=blue,filecolor=blue,urlcolor=blue}
\newcommand{\refs}[1]{{\color{red} insert reference}}
\newcommand{\gcm}[2]{ g cm$^{-2}$}
\newcommand{\gcmm}[3]{ g cm$^{-3}$}

\newcommand{\eq}[1]{Eq.~\ref{#1}}

\usepackage[normalem]{ulem}

\begin{document}


\title{Long-term temperature evolution of neutron stars undergoing episodic accretion outbursts}

\author{L. S. Ootes\inst{1}\thanks{E-mail: l.s.ootes@uva.nl (LSO)}
\and R. Wijnands\inst{1}
\and D. Page \inst{2}}

\institute{Anton Pannekoek Institute for Astronomy, University of Amsterdam, Postbus 94249, 1090 GE Amsterdam, The Netherlands
\and Instituto de Astronom\'{i}a, Universidad Nacional Aut\'{o}noma de M\'{e}xico, Mexico, D.F. 04510, Mexico}

\date{Received date /
Accepted date }

\abstract
{Transient neutron star low-mass X-ray binaries undergo episodes of accretion, alternated with quiescent periods. 
During an accretion outburst, the neutron star heats up due to exothermic accretion-induced processes taking place in the crust. Besides the long-known deep crustal heating of nuclear origin, a likely non-nuclear source of heat, dubbed `shallow heating', is present at lower densities. Most of the accretion-induced heat slowly diffuses into the core on a timescale of years. Over many outburst cycles, a state of equilibrium is reached when the core temperature is high enough that the heating and cooling (photon and neutrino emission) processes are in balance. 
}
{We investigate how stellar characteristics and outburst properties affect the long-term temperature evolution of a transiently accreting neutron star. For the first time the effects of crustal properties are considered, particularly that of shallow heating.}
{Using our code {\tt NSCool}, we tracked the thermal evolution of a neutron star undergoing outbursts over a period of $10^5$~yr. The outburst sequence is based on the regular outbursts observed from the neutron star transient Aql~X-1. For each model we calculated the timescale over which equilibrium was reached and we present these timescales along with the temperature and luminosity parameters of the equilibrium state.}
{We performed several simulations with scaled outburst accretion rates, to vary the amount of heating over the outburst cycles. The results of these models show that the equilibrium core temperature follows a logarithmic decay function with the equilibrium timescale. 
Secondly, we find that shallow heating significantly contributes to the equilibrium state. Increasing its strength raises the equilibrium core temperature. 
We find that if deep crustal heating is replaced by shallow heating alone, the core would still heat up, reaching only a 2\% lower equilibrium core temperature. Deep crustal heating may therefore not be vital to the heating of the core. Additionally, shallow heating can increase the quiescent luminosity to values higher than previously expected. The thermal conductivity in the envelope and crust, including the potentially low-conductivity pasta layer at the bottom of the crust, is unable to significantly alter the long-term internal temperature evolution. Stellar compactness and nucleon pairing in the core change the specific heat and the total neutrino emission rate as a function of temperature, with the consequences for the properties of the equilibrium state depending on the exact details of the assumed pairing models. The presence of direct Urca emission leads to the lowest equilibrium core temperature and the shortest equilibrium timescale.}{}


\maketitle

\section{Introduction}

A neutron star low-mass X-ray binary (LMXB) consists of a neutron star that accretes matter from a low-mass companion (typically with a mass $M\lesssim 1.0\text{ M}_\odot$) via Roche-lobe overflow. While some systems are accreting persistently, most show accretion outbursts alternated with quiescent periods. The latter systems are referred to as transients and the cycles of accretion are often explained as being caused by instabilities in the accretion disc \citep[see e.g. the extensive review by][]{lasota2001}. The accretion outbursts can last from weeks to decades, and during the peak of the outbursts the neutron star can accrete matter at a rate of ${\sim10^{-10}-10^{-8}\text{ M}_\odot\text{ yr}^{-1}}$ \citep[although for some the peak accretion rate is significantly lower, see the discussion in][]{wijnands2006}. In quiescence, no or very little accretion takes place.

When the neutron star is accreting, a layer of material grows on the surface of the star. This freshly accreted material (composed of hydrogen and helium) is fused into heavier elements via stable and unstable thermonuclear processes in the envelope of the neutron star (see e.g. the reviews by \citealt{bildsten1998, Strohmayer2003aa, galloway2017}). As these processes occur at shallow depth (${\rho\sim10^6\text{ g cm}^{-3}}$), most of the energy that is generated is rapidly radiated away from the surface and does not heat the core of the star \citep{brown2000}. Due to the continued buildup of material on the surface, the ashes of the thermonuclear burning are pushed to higher densities, which causes reactions in deeper layers of the neutron star crust. This induces a series of deep crustal heating reactions (electron captures, neutron emission, and pycnonuclear reactions) which release energy deep in the crust \citep{sato1979,haensel1990dch,haensel2008, gupta2007,fantina2018,lau2018}. Additionally, another heat source is inferred to be present from observational data in the outer crust (i.e. at densities lower than the neutron drip, ${\rho\lesssim6\times10^{11}\text{ g cm}^{-3}}$) of accreting neutron stars, referred to as the shallow heat source. A physical explanation for this shallow heat source has not been found \citep[see ][for discussions of the different proposed hypotheses]{deibel2015,waterhouse2016}.

The accretion-induced heating processes cause the crust of the neutron star to be heated up during an outburst \citep{bbr1998,ushomirsky2001}. Although much of the heat flows towards the core, the crust and core get out of thermal equilibrium, because the specific heat of the core is significantly higher than that of the crust and hence the core takes longer to heat up \citep{yakovlev2004,cumming2017}. When the outburst ends, the crust of the neutron star is therefore hotter than the core and the crust will cool down in quiescence in order to restore crust-core thermal equilibrium. Neutron star crust cooling is a phenomenon that has now been observed in nine systems \citep[although in one system crust cooling needs to be confirmed; for an overview of crust-cooling systems see the review by][and references therein]{wijnands2017}. Observational campaigns to follow cooling neutron star transients as soon as they return to quiescence up to decades after the outburst ended are an important tool to gain insight into the physics of neutron stars. The cooling curves that these campaigns provide can be compared with theoretical models to constrain the crustal parameters of a source \citep[e.g.][]{rutledge2002ks,brown2009,page2013,turlione2015,merritt2016,ootes2019}, and even of the core \citep{cumming2017,brown2018}.

The heat generated in a neutron star crust can be emitted either via photon emission from the surface or via various neutrino emission processes from the interior \citep[see e.g.][]{page2006}. However, if the cooling rate is unable to keep up with the rate at which heat from the crust flows towards to core, the neutron star will heat up over time. After many outbursts, the temperature of the core is sufficient to support a high enough cooling rate that can balance the heating and hence an equilibrium state is established \citep{bbr1998,colpi2001}. Once a steady state has been reached, the core temperature will oscillate around the equilibrium core temperature (albeit only by a very small amount) during one accretion cycle (i.e. outburst and following quiescent period), but the average core temperature stays constant. 

The properties of the equilibrium state --  the time needed before equilibrium is reached, the equilibrium core temperature, and the corresponding surface temperature and photon and neutrino luminosities -- depend on the outburst accretion rate and recurrence time, the heating and cooling processes at play, and on the characteristics of the neutron star (e.g. stellar compactness, crustal microphysics, nucleon pairing properties). In this research we investigate how these different properties contribute to the equilibrium state. This study complements previous research by \citet{colpi2001} and \citet{han2017}. Whereas the previous studies were mainly focused on the effects of the core properties, we also set out to investigate if the assumed (micro)physical processes and properties of the crust can have a significant effect on the long-term thermal evolution of an accreting neutron star. Additionally, we have incorporated shallow heating into our model, and used an accretion rate that is based on observed light curves, which introduces more variation in accretion rate during and between outbursts.

\section{Method}

In a simplified model in which the stellar interior is assumed to be always isothermal at a redshifted temperature $\tilde{T}$,
the long-term evolution of a transiently accreting neutron star would be determined by the balance between heating and cooling described by
\begin{equation}
C_V\frac{d\tilde{T}}{dt}=L_\text{H}^\infty(Q_\text{sh},Q_\text{dch})-L_\gamma^\infty -L_{\nu}^\infty \equiv L_\text{H}^\text{eff}
\label{eq:simple}
,\end{equation}
where $C_V$ is the stellar specific heat and $L_\text{H}^\infty$, $ L_\gamma^\infty$, and $L_{\nu}^\infty$ are, respectively, the redshifted heating, photon, and neutrino luminosities. 
The heating luminosity is provided by the amount of deep crustal heating ($Q_\text{dch}$) and shallow heating ($Q_\text{sh}$) that takes place during the outburst. 
The cooling takes place via neutrino emission from the interior (primarily from the core), and photon emission from the surface of the star.
For convenience we also define the total sum of the heat sources and sinks as the `effective' heating luminosity $L_\text{H}^\text{eff}$.
Once equilibrium has been reached, long-term time average values, denoted with $\langle \cdots \rangle$, are then simply related by
\begin{equation}
\langle L_\text{H}^\infty \rangle= \langle L_\gamma^\infty \rangle  + \langle L_{\nu}^\infty \rangle \quad 
\label{eq:equilibrium}
.\end{equation}

To calculate the long-term equilibrium state of a transiently accreting neutron star beyond the above isothermal approximation,
we used our thermal evolution code {\tt NSCool} \citep{page2016}. This code calculates the detailed thermal profile as a function of time for a spherically symmetric neutron star based on the equations for heat transfer and energy conservation (taking into account general relativistic effects). We computed a basic model in which the star is subjected to short accretion outbursts and followed the thermal evolution of the neutron star towards an equilibrium state over a period of $10^5$~yr. We determined the parameters of this equilibrium state (equilibrium core temperature, luminosity, and the time it takes before equilibrium is reached) and then changed the model parameters (describing the outburst accretion rate, recurrence time, stellar compactness, heating processes, crustal microphysics, and assumed pairing gap models) one by one, to determine the effect of each of them on the parameters of the equilibrium state.

\subsection{Accretion-induced heating}

Both the deep crustal heating and the shallow heating are assumed to be directly proportional to the mass accretion rate ($\dot{M}$). In our models, we used the deep crustal heating processes as calculated by \citet{haensel2008}, assuming an initial composition at the top of the crust (defined in our code as a boundary density $\rho_\text{b}=10^8\text{ g cm}^{-3}$) of $^{56}$Fe, left behind by the thermonuclear burning in the envelope layers (which is defined here as $\rho<\rho_\text{b}$). A total amount of $1.9$ MeV per accreted nucleon is released by the electron capture, neutron emission, and pycnonuclear processes. Most of this heat ($\sim 90\%$) is released in the inner crust 
(above the neutron drip at $\rho_\text{drip}=6.2\times10^{11}\text{ g cm}^{-3}$).

Shallow heating is introduced in our models by injecting an amount $Q_\text{sh}$ of heat per accreted nucleon in a region of the outer crust in the density range ${\rho_\text{sh,min}-\rho_\text{sh,max}}$ and is assumed to be uniformly distributed per volume. Because the origin of shallow heating is unknown, we present models in which we varied the strength and depth of the shallow heating. For several crust-cooling sources the amount of shallow heating that is required to explain their observed cooling curves has been constrained to be ${\sim1-2\text{ MeV nucleon}^{-1}}$ \citep[see for example][]{degenaar2011ter5-418,merritt2016,parikh2019mxb,ootes2019}. However, there are also sources that may not require shallow heat \citep[e.g.][]{degenaar2015ter5x3}, and at least one that requires a very strong heat source in the outer crust \citep[${Q_\text{sh}\sim15 \text{ MeV nucleon}^{-1}}$;][]{deibel2015,parikh2017maxi}. Moreover, while for one source the shallow heating was constrained to be consistent between two consecutive outbursts \citep{parikh2019mxb}, for two others the amount and depth of this heating mechanism seems to vary between outbursts \citep{deibel2015,parikh2017maxi,ootes2018}. In the models presented here, we always assume the shallow heating strength and depth to be constant between outbursts.

To simulate accretion outbursts, we did not assume a constant accretion rate during a fixed outburst period, but instead we took into account variations in the outburst history, based on observed light curves \citep{ootes2016}. In this way we simulated accretion outbursts more realistically, although the effect of accretion rate variability on the long-term temperature evolution of a neutron star is minor \citep{cumming2017}.\footnote{We have compared a model in which a time-dependent accretion rate was used with one in which a constant accretion rate was assumed, and we found the difference in equilibrium core temperature for the two models to be negligible (<0.1\%). However, the surface temperature in the crust-cooling phase after an outburst is very sensitive to variation in outburst properties.} Here we used the accretion history of Aql~X-1 observed over the past $\sim20$~yr that we calculated for \citet{ootes2018}. We included here a more recent outburst from 2016 as well, extending the number of outbursts of the sequence to 24 \citep{degenaar2019}. The sequence of outbursts was repeated until a simulation time of  $10^5$~yr was reached (except for some models that required a longer simulation time to reach equilibrium). To be able to repeat the sequence of outbursts, we have taken into account a recurrence time of the last outburst of our sequence of ${0.83 \text{ yr}}$, based on the detection of the start of a new outburst from this source in May 2017 \citep{vlasyuk2017}. Aql~X-1 is a rather useful source from which to use the light curve data, because it shows (contrary to other sources) quite regular outbursts that have been observed over a relatively long period. One might therefore assume the outburst history has been this regular for a long time. On average, the source shows outbursts  with outburst accretion rates on the order  of ${10^{-9}\text{ M}_\odot\text{ yr}^{-1}}$ that last $\sim2.5$ months, and with recurrence times of almost a year. However, amongst the individual accretion episodes, there is still significant variation in outburst properties \citep[e.g.][]{gungor2017,ootes2018}. Despite the fact that these outburst characteristics make Aql X-1 a very useful source for our studies, we note that it cannot necessarily be regarded as a typical neutron star low-mass X-ray binary transient, as most other sources show more infrequent outburst behaviour \citep[e.g. see the overviews by][]{chen1997,yan2015}. Additionally, Aql~X-1 has a relatively long orbital period of 19 hours \citep{chevalier1991,welsh2000} compared to the other neutron star X-ray transients \citep[see e.g. the lists provided in][]{wu2010,lin2019}.

\subsection{Heat transport, neutrino emission, and superfluidity}

Heat that is generated in the crust during an outburst spreads towards the core of the star and towards the surface. The thermal conductivity inside the crust is set by electrons scattering off lattice impurities, phonons/ions, and other electrons \citep{yakovlev1980,gnedin2001,shternin2006}. The first observational crust-cooling studies revealed that the crust of an accreting neutron star cooled fast in quiescence, indicating that the thermal conductivity of this region is high \citep{wijnands2002,wijnands2004,cackett2006,shternin2007}. This observational result was initially unexpected from a theoretical perspective. To achieve such high thermal conductivity, the crust would need to be structured in a highly ordered lattice, rather than it being amorphous \citep{horowitz2009}. The innermost region of the crust may host a so-called nuclear pasta region \citep{ravenhall1983,hashimoto1984}. In this region, the density approaches the nuclear density. which causes a competition between the attractive nuclear force and the repulsive Coulomb force. As a consequence, matter can take on several pasta-like shapes \citep[see for a recent review][]{caplan2017}. Such a region is predicted to have a low thermal conductivity, and can hence slow down the heat flow towards the core \citep[e.g.][]{horowitz2015}. In this study, we consider the effect of the thermal conductivity of the crust on the temperature evolution of the neutron star by changing the impurity factor $Q_\text{imp}$ (which quantifies how impure the lattice structure is) of the crust, and investigate if a low-conductivity pasta region can have a significant effect on the equilibrium core temperature of a transiently accreting neutron star.

The (redshifted) photon luminosity of the neutron star is set by 
\begin{equation}
L_\gamma^\infty=4\pi (R^\infty)^2 \sigma (T_\text{eff}^\infty)^4 \label{eq:photonlum}
,\end{equation}
where $R^\infty$ is the radius of the neutron star at infinity, $\sigma$ the Stefan-Boltzmann constant, and $T_\text{eff}^\infty$ is the redshifted effective temperature. Within {\tt NSCool}, the internal temperature profile is calculated up to the boundary density $\rho_\text{b}$. We refer to the layers at $\rho<\rho_\text{b}$ as the envelope. This region acts as isolation between the hot crust and the surface. Hence, there is a strong temperature gradient across this thin (approximately tens of meters) layer. Detailed calculations have been carried out to derive $T_\text{b}-T_\text{eff}^\infty$ relations to calculate the effective temperature for a particular boundary temperature, $T_\text{b}(\rho_\text{b})$ \citep[e.g.][]{gudmundsson1983}. For accreting neutron stars, the envelope strongly varies in composition as a function of time during outburst due to thermonuclear burning (the details of which depend on the accretion rate and composition of accreted material) and the accumulation of newly accreted material \citep[e.g.][]{woosley2004}. As a consequence, the composition of the envelope in quiescence can differ from outburst to outburst \citep{brown2002}, depending on when the last X-ray burst occurred and the type of the burst. The envelope consists of a layer of accreted material (hydrogen and helium) on top of layers containing heavier elements, which are products of thermonuclear burning. The lighter elements in the envelope have a higher thermal conductivity than heavier elements \citep[e.g.][]{potekhin1999}. Consequently, for a specific temperature $T_\text{b}$ at the bottom of the envelope, the effective temperature will be higher for an envelope that contains a thick light element layer, compared to an envelope containing more heavy elements \citep{potekhin1997}. We used here the $T_\text{b}-T_\text{eff}^\infty$ relations as outlined in \citet{ootes2018}, which depend on the light element column depth $y_\text{L}$. The light element layer is assumed to consist of helium and carbon (helium has a stronger effect on the $T_\text{b}-T_\text{eff}^\infty$ relation than hydrogen). During our simulations we kept the envelope composition constant between different outburst, although we used different light element column depths for two simulations to investigate the effect of variation in the thermal conductivity of the envelope on the equilibrium state. 

Most of the heat generated in the crust will flow towards the core. The core of a neutron star has a high thermal conductivity \citep[][]{baiko2001}, such that (contrary to the crust) the core is always practically isothermal. The timescale over which the stellar core heats up depends on the specific heat, which is highly uncertain because it depends strongly on the state of the matter inside the core. If the nucleons are paired into Cooper pairs, forming a superfluid (neutron pairing) or superconductor (proton pairing), the specific heat can be either reduced or enhanced, depending on the temperature \citep[see e.g. Fig. 13 in][]{page2004}. Several spin-angular momentum channels are possible in which pairing can occur in the core, mostly likely to be 
$^3$P$_2$ for neutrons and $^1$S$_0$ for protons, but with large uncertainties about the value of the corresponding,
density-dependent critical temperatures $T_\text{crit}(\rho)$. In the temperature evolution models presented here we used several pairing gap models and also investigated the possibility that the nucleons are unpaired. We refer to \citet{page2014} for an extensive review on superfluidity in neutron stars.

Cooling via neutrino emission can take place throughout the entire neutron star via various neutrino emission processes \citep[see e.g. the review by][]{yakovlev2001}. These processes become more efficient at higher temperatures and the strongest neutrino emission processes take place in the neutron star core. The various neutrino processes are generally divided into fast and slow neutrino emission processes. The slow neutrino emission processes include modified Urca (mUrca), Bremsstrahlung, and Cooper Pair Breaking and Formation (PBF). Fast neutrino emission in neutron stars can be generated by the direct Urca process \citep[dUrca,][]{Boguta:1981aa,lattimer1991}. The emissivity of this process is several orders of magnitude stronger than that of the slow neutrino emission processes, and hence direct Urca results in much stronger cooling \citep[e.g.][]{page1992}, but is expected to occur only in the core of massive neutron stars (see e.g. \citealt{brown2018} for a recent example). Just like the specific heat, the strength of the neutrino emission in a neutron star depends on the presence of nucleon pairing and can be reduced or enhanced depending on the temperature.

\subsection{Calculating the equilibrium state}

For our models, we assumed the equation of state (EOS) for the core (${\rho>1.5\times10^{14}\text{ g cm}^{-3}}$) from \citet{akmal1998}. This equation of state assumes only `regular' neutron star material, and we do not consider the effects of exotic matter. For the crust, we used the accreted crust model from \citet{haensel2008}. We thus assumed that the neutron star has accreted for a period long enough to fully replace its original, catalysed crust \citep[see][for a discussion on partly-accreted crusts]{wijnands2013}. Furthermore, we assumed an initial (redshifted) core temperature of ${\tilde{T}_0=10^6 \text{ K}}$ at the start of the simulation, and repeated the sequence of accretion outbursts (24 outbursts over a duration of ${\sim20\text{ yr}}$) until the simulation reached a time of $10^5$~yr. The heat flow towards the core is very different between outburst and quiescence, because of the very different temperature profiles in the crust in these two phases. Hence, once the star has reached an equilibrium between heating and cooling, the core temperature will still oscillate around the equilibrium temperature over an outburst cycle, even though the variation is very small \citep[see Fig.~2 in][]{colpi2001}. We therefore collected the temperatures and luminosities at the end of each sequence of 24 outbursts, at the last time step before the start of the first outburst of the next sequence. This is (except for the models in which the recurrence time is increased) during the crust-cooling phase, because of the short recurrence time of the outbursts in Aql X-1 \citep{ootes2018}. Consequently, the surface temperature and photon luminosity are different compared to what these values would be if the star were in crust-core equilibrium, but the core temperature and neutrino luminosity are not affected. We defined the time at which equilibrium is reached from this data as the time when the temperature reaches $99.9\%$ of the maximum temperature.

\section{Results and discussion}
 \begin{figure*}
  \includegraphics[width=0.5\textwidth]{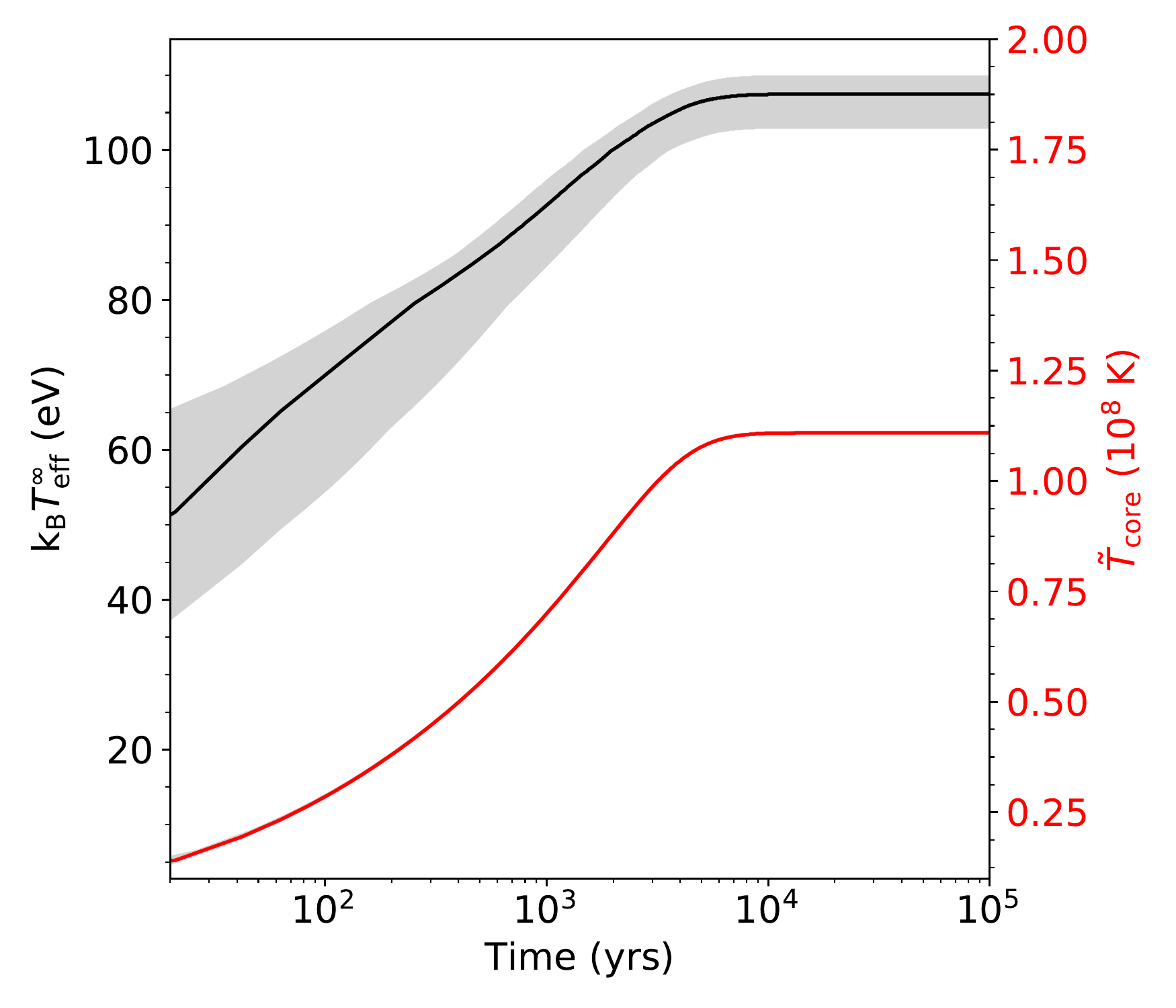}\includegraphics[width=0.5\textwidth]{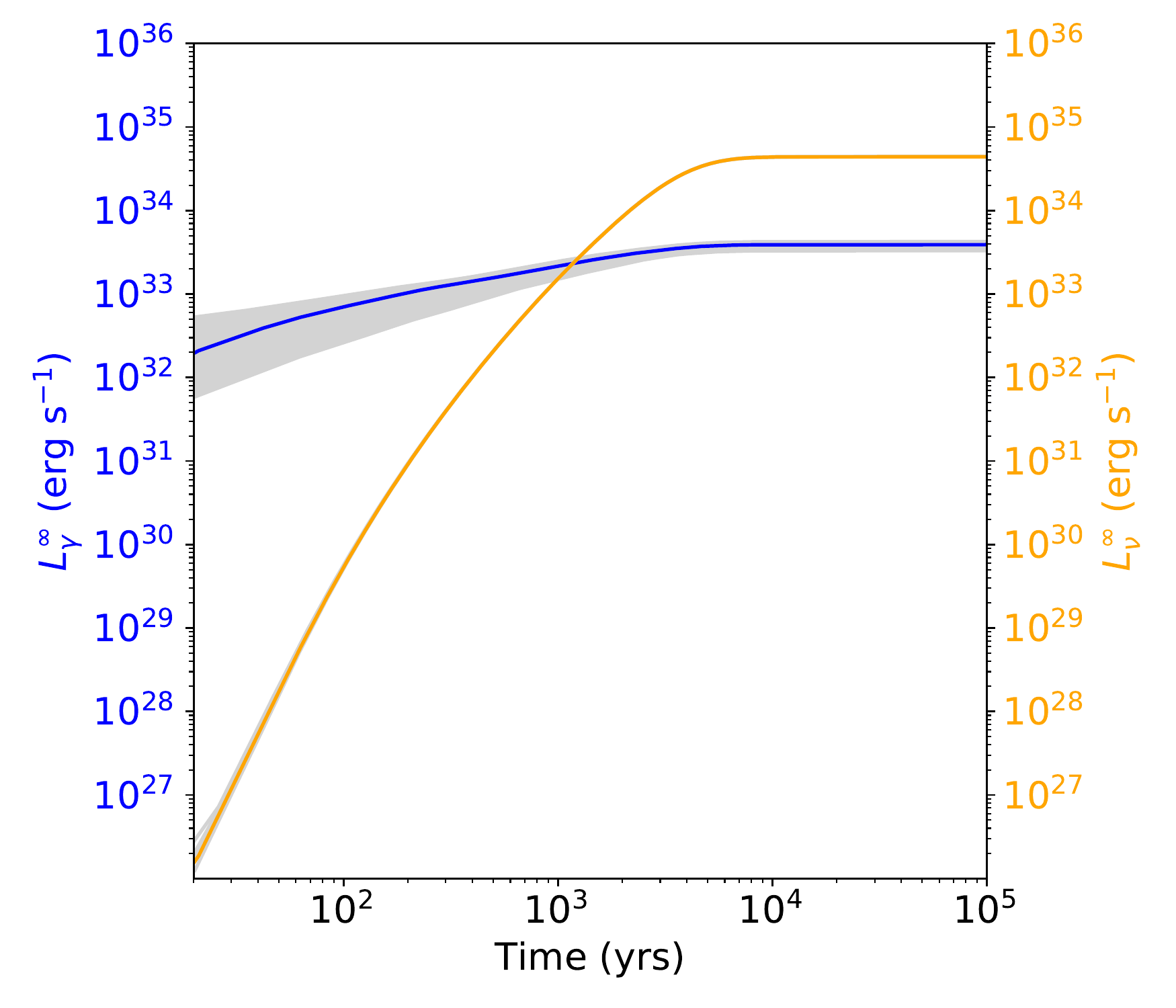}
     \caption{Results of the basic model. Left: Evolution of the effective temperature (black) and core temperature (red) as function of time for our basic model. Right: Evolution of the photon luminosity (blue) and neutrino luminosity (orange). All values are redshifted. Plotted are values at the end of the last of the 24 outburst that make up one sequence, at the last step of the quiescent period. The plot thus excludes the variations during outbursts and between the 24 different outbursts of one sequence. The grey band indicates how different the parameters would be if they were taken at the end of the quiescent period after one of the 23 other outbursts in the sequence. }
     \label{fig:basic}
\end{figure*}

\begin{table*}[]
\caption{Derived properties of the equilibrium state for the different models. In the presented models a sequence of 24 outbursts is repeated. To eliminate the variation in the temperatures and luminosities due to the oscillation around the equilibrium state over the outbursts cycles in the sequence, in each model the parameter values are tracked at the last time step in quiescence at the end of the full sequence (24 outbursts). The equilibrium temperature is defined as the time when the core temperature reaches 99.9\% of the maximum core temperature (from the core temperatures at the tracked times).}
\label{tab:eqtemps}      
\centering 
\begin{tabular}{ccccccl}
\hline\hline
  & $t_\mathrm{eq}$ & $\tilde{T}_{\mathrm{core, eq}}$ & $T_\mathrm{eff,eq}^\infty$ & $L_\mathrm{\gamma,eq}^\infty$ & $L_{\nu,\mathrm{eq}}^\infty$    & Description       \\
   &         ($10^3$~yr)                        & ($10^8$ K)                             & (eV)                                          & ($10^{33} $ erg s$^{-1}$)       & 
   ($10^{33}$ erg s$^{-1}$) &\\
   \hline
1       &        9.7            &       1.109   &        107    &       3.90    &        44.0 &   basic  \\
2       &        42.0   &        0.770  &         86    &       1.59    &        \;\;3.1 &        $t_\mathrm{rec}*10$    \\
3       &         112.7 &       0.526   &         73    &       0.82    &        \;\;0.2 &        $f_\text{a}=0.02$      \\
4       &        40.2   &        0.773  &         87    &       1.67    &         \;\;3.2 &       $f_\text{a}=0.1$       \\
5       &        14.1   &       1.002   &        100    &       2.98    &        21.2 &   $f_\text{a}=0.5$       \\
6       &        6.1            &       1.222   &        117    &       5.57    &        89.0 &   $f_\text{a}=2.0$       \\
7       &        13.0   &       1.026   &         99    &       2.78    &        25.2 &   No shallow heating ($Q_\mathrm{sh}=0)$ \\
8       &        16.8   &        0.972  &        102    &       3.12    &        17.0 &   No deep crustal heating ($Q_\mathrm{dch}=0)$   \\
9       &        10.1   &       1.098   &        106    &       3.73    &        41.0 &  $\rho_\mathrm{sh}=(2-10)        \times10^8$ g cm$^{-3}$ \\
10      &        9.4            &       1.115   &        107    &       3.88    &        45.8 &   $\rho_\mathrm{sh}=(1-5)\times10^{10}$ g cm$^{-3}$      \\
11      &        6.1            &       1.213   &        122    &       6.49    &        84.5 &   $Q_\mathrm{sh}=5.0$ MeV nuc$^{-1}$     \\
12      &        4.3            &       1.275   &        111    &       3.86    &        48.8 &   $M=1.2\text{ M}_\odot$ \\
13      &        16.2   &       1.007   &        106    &       4.23    &        37.9 &   $M=2.0\text{ M}_\odot$ (minimal cooling paradigm)      \\
14      &        8.8            &       1.117   &         90    &       1.90    &        46.6 &   $y_\mathrm{L}=10^{6}$ g cm$^{-2}$      \\
15      &        9.9            &       1.096   &        121    &       6.32    &        40.5 &   $y_\mathrm{L}=10^{11}$ g cm$^{-2}$     \\
16      &        9.8            &       1.104   &        114    &       4.93    &        42.6 &   $Q_\mathrm{imp}=30$    \\
17      &        9.9            &       1.094   &        125    &       7.06    &        40.0 &   $Q_\mathrm{imp}=100$   \\
18      &        9.1            &       1.108   &        108    &       4.00    &        43.9 &   $Q_\mathrm{imp,p}=30$  \\
19      &        9.2            &       1.107   &        110    &       4.28    &        43.6 &   $Q_\mathrm{imp,p}=100$ \\
20      &        10.1   &       1.108   &        107    &       3.90    &        44.0 &   neutron $^1\text{S}_0$ gap model `CCDK'        \\
21      &        10.5   &       1.108   &        107    &       3.90    &        44.1 &   no neutron $^1\text{S}_0$ pairing      \\
22      &        11.4   &       1.837   &        129    &       8.08    &        39.4 &   neutron $^3\text{P}_2$ gap model `c'   \\
23      &        37.6   &       1.732   &        126    &       7.47    &        40.2 &   no neutron $^3\text{P}_2$ pairing      \\
24      &        7.2            &       1.109   &        107    &       3.90    &        44.0 &   proton $^1\text{S}_0$ gap model `CCDK' \\
25      &        10.4   &       1.109   &        107    &       3.90    &        44.0 &   no proton $^1\text{S}_0$ pairing       \\
26      &        0.3            &        0.194  &         49    &       0.19         &        41.4 &  $M=2.0\text{ M}_\odot$ with dUrca      \\
\hline
\end{tabular}
\end{table*}

\subsection{Basic model}
For our basic model (model 1), we took a neutron star with a mass ${M=1.6 \text{ M}_\odot}$, which for the EOS that we used gives a stellar radius of 11.5 km. The average outburst accretion rate is $\langle\dot{M}_\text{ob}\rangle{\sim1.4\times10^{-9}\text{ M}_\odot\text{ yr}^{-1}}$, which results in a long-term time-averaged accretion rate of $\langle\dot{M}\rangle{\sim 3.2\times 10^{-10}\text{ M}_\odot\text{ yr}^{-1}}$. In addition to the heat released by deep crustal heating during outburst, we set the shallow heating strength to $Q_\text{sh}=1.5 \text{ MeV nucleon}^{-1}$ released over a density range $\Delta\,  \rho_\text{sh}=[1-5]\times10^{9}\text{ g cm}^{-3}$. We assumed that the crust has a high thermal conductivity and set $Q_\text{imp}=1$, throughout the whole crust. For the envelope we take in our basic model ${y_\text{L}=10^9 \text{ g cm}^{-2}}$. In regards to the pairing of nucleons, we assumed the critical temperatures from the neutron $^1\text{S}_0$ superfluid gap calculated by \citet{schwenk2003}. This model has a maximum critical temperature ${T_\text{crit,max}\sim5\times10^9\text{ K}}$ and the gap continues up to the crust-core boundary such that the neutrons can be superfluid in the entire bottom of the crust. For the core pairing, we assumed the neutron $^3\text{P}_2$ gap `a' from \citet{page2004}, which is based on the results of \citet{baldo1998}, with $T_\text{crit,max}=10^9\text{ K}$, and the proton $^1\text{S}_0$ gap from \citet{amundsen1985}, with $T_\text{crit,max}\sim2.5\times10^9\text{ K}$. For this neutron star the central density is not high enough for direct Urca to occur, and hence only the neutrino emission processes from the minimal cooling scenario \citep{page2004} are allowed.

Figure~\ref{fig:basic} shows the results for the temperature evolution (left panel, with the black curve the effective temperature and the red curve the redshifted core temperature) and the luminosities (right panel, with the blue curve the photon luminosity and the orange curve the neutrino luminosity). During one sequence of 24 outbursts, there is significant variation in the temperatures and luminosities. Because we are interested in the equilibrium state that is reached over many cycles of the outburst sequence, we plot in Fig.~\ref{fig:basic} the values for the temperatures and luminosities once per outburst sequence. We have chosen to plot the parameter values at the end of the quiescent period after the last outburst of our sequence (i.e. after every $\sim20$~yr, at the last time step before the onset of the first outburst of the next sequence). However, in \citet{ootes2018} we found that the recurrence times of the outbursts from Aql~X-1 are so short, that -- under the assumed heating behaviour and neutron star properties -- at the onset of a new outburst some thermal energy remains in the crust of the neutron star, meaning that the temperature profile is not such as would be the case for crust-core equilibrium. Since the outburst duration, outburst accretion rate, and recurrence time is different for each outburst in the sequence, the temperature profile is different at the onset of each new outburst. Consequently, in particular the effective temperature would be different if it were `measured' after a different outburst in the sequence. The grey areas in Fig.~\ref{fig:basic} indicate the spread on the respective curves that is introduced by taking the value after different outbursts. Although the choice of the outburst after which the properties defining the equilibrium state are `measured' affects the inferred effective temperature shown in Fig.~\ref{fig:basic}, the core temperature evolution is not affected by this choice. Therefore, from here onwards we only consider the values taken at the end of each last outburst of the outburst sequence. 

In Table \ref{tab:eqtemps} the properties of the equilibrium state are presented. For the basic model, we find that after $9.7\times10^3$~yr the heating during the outbursts is in equilibrium with the cooling. The core temperature at equilibrium is $\tilde{T}_\text{core,eq}=1.109\times10^8\text{ K}$. After $\sim10^3$~yr the neutrino luminosity exceeds the photon luminosity in this model. When equilibrium is reached, the neutrino luminosity is a factor of approximately ten higher than the photon luminosity, and it is thus the neutrino luminosity that sets the equilibrium properties. 

\subsection{Assuming different outburst properties}

\subsubsection{Recurrence time}

First we investigate the effect on the outcome of our simulations when assuming different outburst properties. {Aql~X-1} has outbursts with relatively short recurrence times of ${\sim0.9\text{ yr}}$ (on average). In model 2 we increased the average recurrence time by a factor of 10 to 9.0~yr, while keeping the outburst strength and duration the same compared to model 1. Such a recurrence time in combination with the short outburst duration of Aql~X-1, makes the outburst behaviour of model 2 more similar to other neutron star LMXB transients that recur less often than Aql~X-1 \citep{chen1997}. Therefore, this model might be more representative for the average transient neutron star LMXB.

As expected, increasing the recurrence time has a significant effect on the long-term temperature evolution. Although the amount of heat that enters the star per outburst cycle remains the same compared to model 1, the source has a longer period over which it can cool, such that the time-averaged heating is lower than for model 1. Hence in its equilibrium state the core temperature is significantly lower (${\tilde{T}_\text{core,eq}=0.77\times10^8\text{ K}}$) compared to model 1 (see Table \ref{tab:eqtemps} and Fig.~\ref{fig:trec}) and the equilibrium photon luminosity is now about 50\% of the neutrino one.
Furthermore, in model 2 it takes about four times longer before the equilibrium state is reached (${t_\text{eq}=4.20\times10^4\text{ yr}}$). From \eq{eq:equilibrium}, if heating were balanced predominantly by neutrino losses with ${L_\nu \propto \tilde{T}^8}$ and $\langle L_\text{H} \rangle$ is reduced by a factor of ten as we imposed here, then $\tilde{T}_\text{core,eq}$ should be reduced by ${10^{1/8} \approx 1.33}$, that is, from $1.1\times 10^8$~K in model 1 down to ${0.83 \times 10^8}$~K. However, in model 2, $L_\gamma$ contributes significantly and hence a slightly lower ${\tilde{T}_\text{core,eq}=0.77\times10^8\text{ K}}$ is obtained.

One can estimate the thermal equilibration time $\tau_\text{th}$ like \citet{wijnands2013},
\begin{equation}
\tau_\text{th} \approx \frac{C \tilde{T}_\text{core,eq}^2}{2 \langle L_\text{H} \rangle}
\label{eq:tau}
,\end{equation}
under the assumption that the specific heat is $C_V = C \tilde{T}_\text{core}$, with $C$ a temperature-independent constant, as for a normal degenerate Fermion system. In our case we found that $\tilde{T}_\text{core,eq}$ is reduced by a factor of $0.7$ from model 1 to model 2, while $\langle L_\text{H} \rangle$ is suppressed by a factor of ten, so we expect that $\tau_\text{th}$ is increased by a factor of 4.8 from \eq{eq:tau}. This is in good agreement with the numerically found factor $4.20\times10^4\text{ yr}/9.7\times10^3\text{ yr} \approx 4.3$.

 \begin{figure}
  \includegraphics[width=0.5\textwidth]{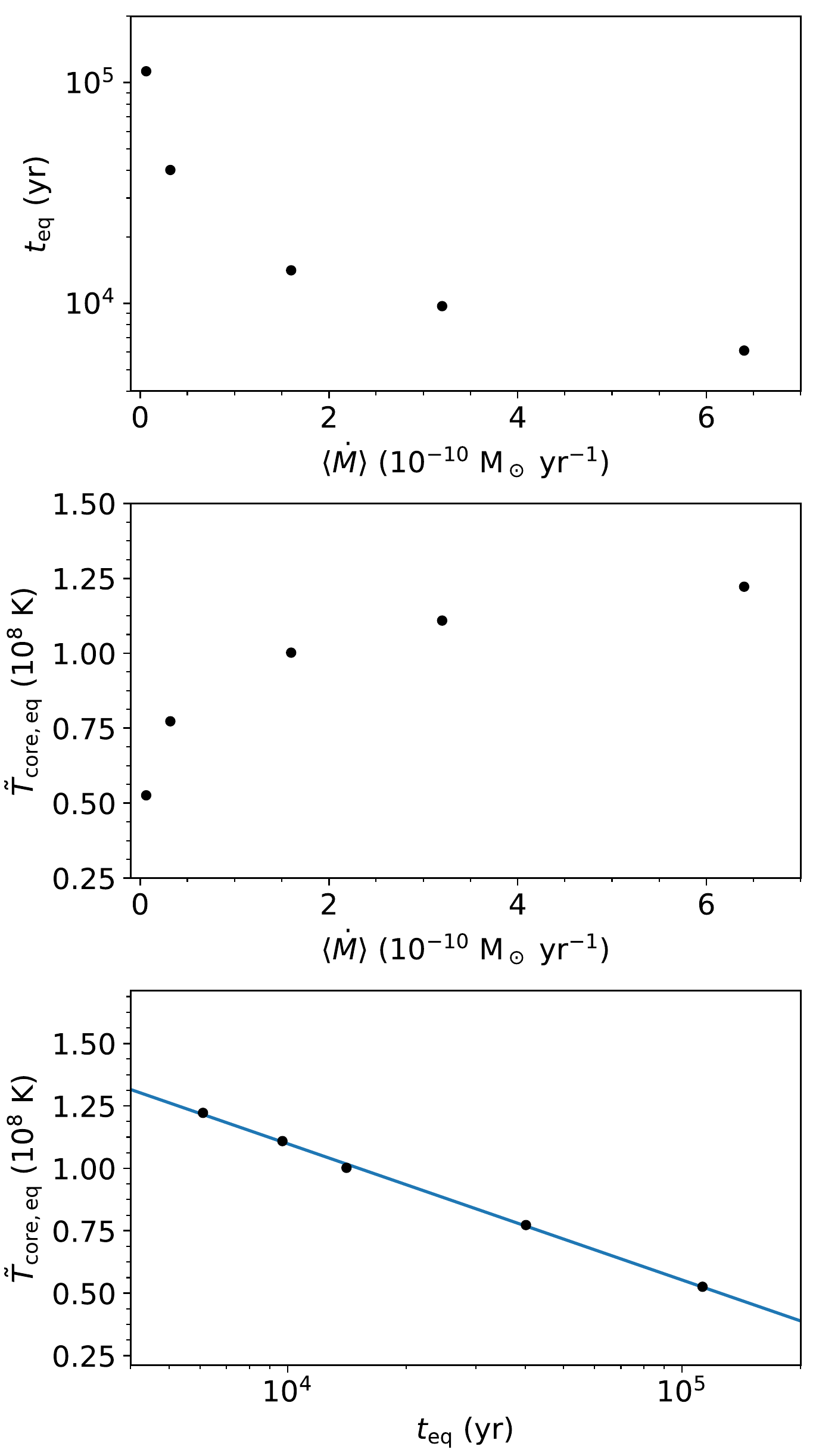}
     \caption{Relation between the equilibrium core temperature, equilibrium timescale, and time-averaged accretion rate for models 1, 3, 4, 5, and 6.}
     \label{fig:mdotrelation}
\end{figure}

\subsubsection{Average outburst accretion rate}

Next, we investigate the effect of assuming different average outburst accretion rates on the thermal equilibrium. In the basic model, the outbursts have an average outburst accretion rate of ${\langle\dot{M}_\text{ob}\rangle\sim1.4\times10^{-9}\text{ M}_\odot\text{ yr}^{-1}}$. In models 3--6 we multiplied the accretion rates by factors of ${f_\text{a}=0.02}$, ${f_\text{a}=0.1}$, ${f_\text{a}=0.5}$, and ${f_\text{a}=2.0}$ respectively, while keeping the other model properties the same as for model 1. 
As a result, the time-averaged accretion rates for models 3--6 become respectively ${\langle\dot{M}\rangle\sim{6.4\times10^{-12}\text{ M}_\odot\text{ yr}^{-1}}}$,  ${\langle\dot{M}\rangle\sim3.2\times10^{-11}\text{ M}_\odot\text{ yr}^{-1}}$, ${\langle\dot{M}\rangle\sim1.6\times10^{-10}\text{ M}_\odot\text{ yr}^{-1}}$, and ${\langle\dot{M}\rangle\sim6.4\times10^{-10}\text{ M}_\odot\text{ yr}^{-1}}$. 
For model 3, the total time of the simulation was extended to $10^6$~yr,   in order to determine the properties of the equilibrium state, because for such a low average accretion rate the core takes longer to heat up.

The results of models 3--6 are shown in comparison with model 1 in Fig.~\ref{fig:mdot} and Table \ref{tab:eqtemps}. For models 3--6, the results are as expected. Decreasing the average outburst accretion rate decreases the equilibrium core temperature (${\tilde{T}_\text{core,eq}=0.526\times10^8\text{ K}}$ for model 3, ${\tilde{T}_\text{core,eq}=0.773\times10^8\text{ K}}$ for model 4, and ${\tilde{T}_\text{core,eq}=1.002\times10^8\text{ K}}$ for model 5) and increases the time in which the equilibrium state is reached (${t_\text{eq}=112.7\times10^3\text{ yr}}$ for model 3, ${t_\text{eq}=40.2\times10^3\text{ yr}}$ for model 4, and ${t_\text{eq}=14.1\times10^3\text{ yr}}$ for model 5), and vice versa for model 6 in which the outburst accretion rate is increased by a factor of two (${\tilde{T}_\text{core,eq}=1.222\times10^8\text{ K}}$ and ${t_\text{eq}=6.1\times10^3\text{ yr}}$). In model 3 the equilibrium core temperature is low enough that the photon luminosity is dominant over the neutrino luminosity (see right panel of Fig. \ref{fig:mdot}).
In these cases, again, the simple estimates of $\tau_\text{th}$ from \eq{eq:tau} are in good agreement with the numerically found $t_\text{eq}$.

In Fig.~\ref{fig:mdotrelation}, we plot the relation between the time-averaged accretion rate, equilibrium core temperature, and equilibrium time that we obtained for models 1, 3, 4, 5, and 6. The equilibrium core temperature decreases with time-averaged accretion rate, while the equilibrium timescale increases with the time-averaged accretion rate. The equilibrium core temperature as a function of the equilibrium time for these models can be fitted with a logarithmic decay function.

Model 4 has the same time-averaged accretion rate as model 2 and naturally gives very similar results for the equilibrium values, but with nevertheless interesting differences (see Table \ref{tab:eqtemps}). 
Although the time-averaged accretion rate is the same, the models differ in how these are achieved. Model 4 has a short recurrence time, with low outburst accretion rates, and model 2 has long recurrence times with high outburst accretion rates. As a result, during the outburst and the subsequent cooling phase the star in model 2 is hotter and thus energy losses, both $L_\gamma$ and $L_\nu$, are somewhat larger than in model 4: the effective heating luminosity $L_\text{H}^\text{eff}$ is hence slightly lower, explaining the slightly lower equilibrium temperature and the slightly longer equilibrium time. However, we emphasise that these differences are only minor.

\subsection{Changing the heating properties}

In our basic model (model 1), heating during the outbursts was assumed to be caused by deep crustal heating (with a total strength of $1.9\text{ MeV nucleon}^{-1}$) and shallow heating (with a strength of $1.5\text{ MeV nucleon}^{-1}$, released in the outer crust at a depth between ${[1-5]\times10^9\text{ g cm}^{-3}}$). In Fig.~\ref{fig:heating} we present model 7 in which we assumed that only deep crustal heating takes place (and no shallow heating), and model 8, in which only shallow heating was activated. The properties of the heating mechanism that remains active in models 7 and 8 have not been changed compared to model 1. The results show that for model 1 the contribution of the deep crustal heating process is most significant to the heating of the neutron star to an equilibrium state, but that shallow heating is significant as well. When shallow heating is deactivated (model 7), the core temperature evolution curve is 7\% lower compared to model 1 once equilibrium is reached (${\tilde{T}_\text{core,eq}=1.026\times10^8\text{ K}}$, see Table \ref{tab:eqtemps}). For model 8 (no deep crustal heating), the equilibrium core temperature is 12\% lower than for model 1 ($\tilde{T}_\text{core,eq}=9.72\times10^7\text{ K}$). The difference between models 7 and 8 is due to the difference in heating strength assumed for the two heating mechanisms ($1.9\text{ MeV nucleon}^{-1}$ for deep crustal heating versus $1.5\text{ MeV nucleon}^{-1}$ for shallow heating), and due to the difference in depth at which the heat is released (primarily in the inner crust for deep crustal heating and in the outer crust for shallow heating).

Figure~\ref{fig:heating} shows that model 7 has a higher core temperature than model 8 throughout the simulation, but that the surface temperatures of these two models are inverted; the surface temperature is higher for model 8 than for model 7. The reason for this trend is that in model 8 (only shallow heating), due to the shallow depth of the heat source, a larger fraction of the heat flows to the surface compared to model 7 (only deep crustal heating). This causes the surface temperature to be higher throughout the crust-cooling phase in quiescence (when the surface temperatures are `measured' from the models). 

In models 9 and 10 the effect of varying the depth of heat release is investigated. In these two models both shallow heating and deep crustal heating are assumed to take place. We kept the shallow heating strength equal to that of model 1, but in model 9 we placed it at a shallower depth (in the density range ${\Delta\,  \rho_\text{sh}=[2-10]\times10^8\text{ g cm}^{-3}}$), and in model 10 we placed it deeper (in the density range ${\Delta\,  \rho_\text{sh}=[1-5]\times10^{10}\text{ g cm}^{-3}}$). Figure~\ref{fig:rhosh} shows that variations in the shallow heating depth have a marginal effect on the equilibrium parameters. The equilibrium core temperature is highest for the model that has the deepest shallow heating (model 10), indicating that a larger fraction of the heat flows towards the core for deeper heating resulting in a slightly larger $L_\text{H}^\text{eff}$. However, the equilibrium core temperature for model 9 (shallower heating) is only $\sim1\%$ lower compared to model 1, and for the model that assumes deeper shallow heating (model 10) the core temperature in equilibrium is  $\sim0.5\%$ higher than for the basic model (see Table~\ref{tab:eqtemps}). 

The evolution of the effective temperature shown in the left panel of  Fig.~\ref{fig:rhosh} is compatible with the evolution of $\tilde{T}_\text{core}$, but with a twist. The effective temperature $T_\text{eff}^\infty$ is determined by the temperature in the outer layers of the crust and the values we display in the figure  are `measured' at the end of the last cooling phase (which is 236~days after the outburst end) in each sequence of 24 outbursts when the crust is somewhat, but never completely, relaxed from the accretion heating. When compared to model 1, in model 10 a larger part of $Q_\text{sh}$ flowed inward into the core resulting in a cooler crust and a lower $T_\text{eff}^\infty$ at the late time when we measure it, while in model 9 an even larger part of $Q_\text{sh}$ flowed outward, resulting in a higher $T_\text{eff}^\infty$ at early time during the post-accretion cooling phase (not shown in the figures) but in a cooler crust and lower $T_\text{eff}^\infty$ at late time, when we measure it for the figures. As time passes and the star reaches the equilibrium state, the difference in $T_\text{eff}^\infty$ between models 1 and 10 becomes insignificant while in model 9 the surface losses are large enough that $T_\text{eff}^\infty$ remains lower. The differences are nevertheless very small and smaller than the statistical error in the observed $T_\text{eff}^\infty$.

Finally, Fig.~\ref{fig:Qsh} shows the effect of varying the shallow heating strength on the equilibrium state of the neutron star. In this figure, model 1 is compared to model 7 in which the shallow heating has been deactivated, and model 11 in which the shallow heating strength has been increased to $Q_\text{sh}=5\text{ MeV nucleon}^{-1}$. Comparing to models 9 and 10, the effect of variation in shallow heating strength is much more significant than when the depth of the shallow heating was varied. For stronger shallow heating ($Q_\text{sh}=5\text{ MeV nucleon}^{-1}$) the equilibrium core temperature is $9\%$ higher (${{T}_\text{core,eq}^\infty=1.213\times10^8\text{ K}}$) than for model 1. The effect is similar compared to model 6 (see Fig.~\ref{fig:mdot}) in which the accretion rate was increased by a factor of two. In both models an effectively similar amount of heat is released in the crust during outbursts (i.e. $6.9\text{ MeV nucleon}^{-1}$ in total from shallow heating and deep crustal heating for model 11 versus $3.4\text{ MeV nucleon}^{-1}$ but at a rate that is two times higher for model 6), but because for model 11 the heating is only increased at shallow depths (while in model 6 more heat is released from deep crustal heating as well), the equilibrium core temperature in model 11 is slightly lower ($\sim0.7\%$, see Table~\ref{tab:eqtemps}) compared to model 6.

From these models it becomes clear that shallow heating can play an important role in the long-term heating of transiently accreting neutron stars (assuming that shallow heating is active during all outbursts). The strength of the shallow heating has a stronger effect on the equilibrium core temperature than the depth of the shallow heating. Even when shallow heating is placed just below the envelope, most of the heat released during outbursts will flow towards the core rather than the surface. As shown in Fig.~\ref{fig:heating}, if deep crustal heating is not active, shallow heating heats up the core of the neutron star. The equilibrium core temperature is $5\%$ lower than if only deep crustal heating is assumed to take place and this difference would be smaller ($2\%$) if the same heating strength were assumed for the two heating mechanisms. Moreover, for shallow heating strengths in excess of the heating strength from deep crustal heating, the contribution of shallow heating to the equilibrium state can become dominant. Deep crustal heating is therefore not essential to long-term heating of transiently accreting neutron stars if significant shallow heating takes place.

 \begin{figure*}
  \includegraphics[width=0.5\textwidth]{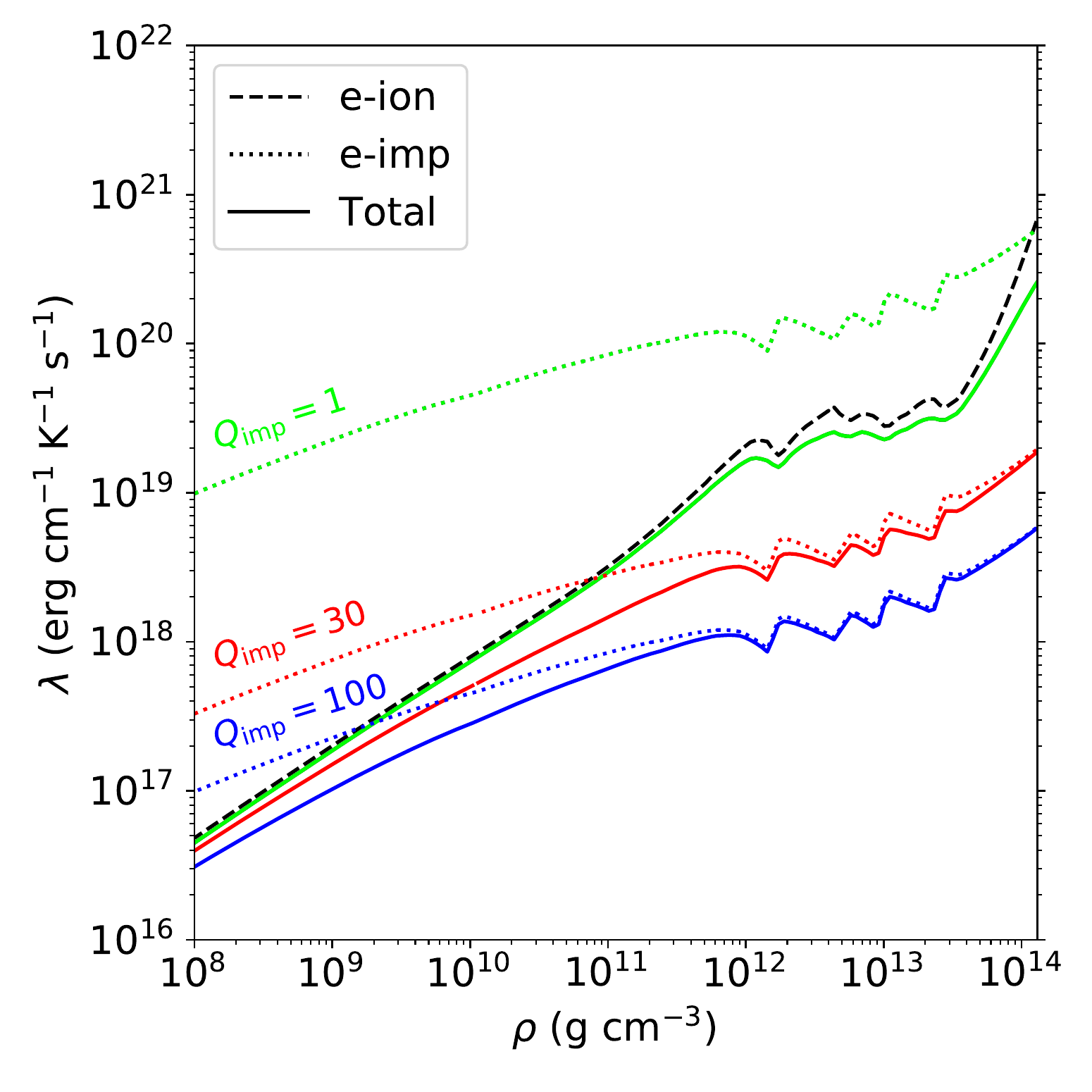}\includegraphics[width=0.5\textwidth]{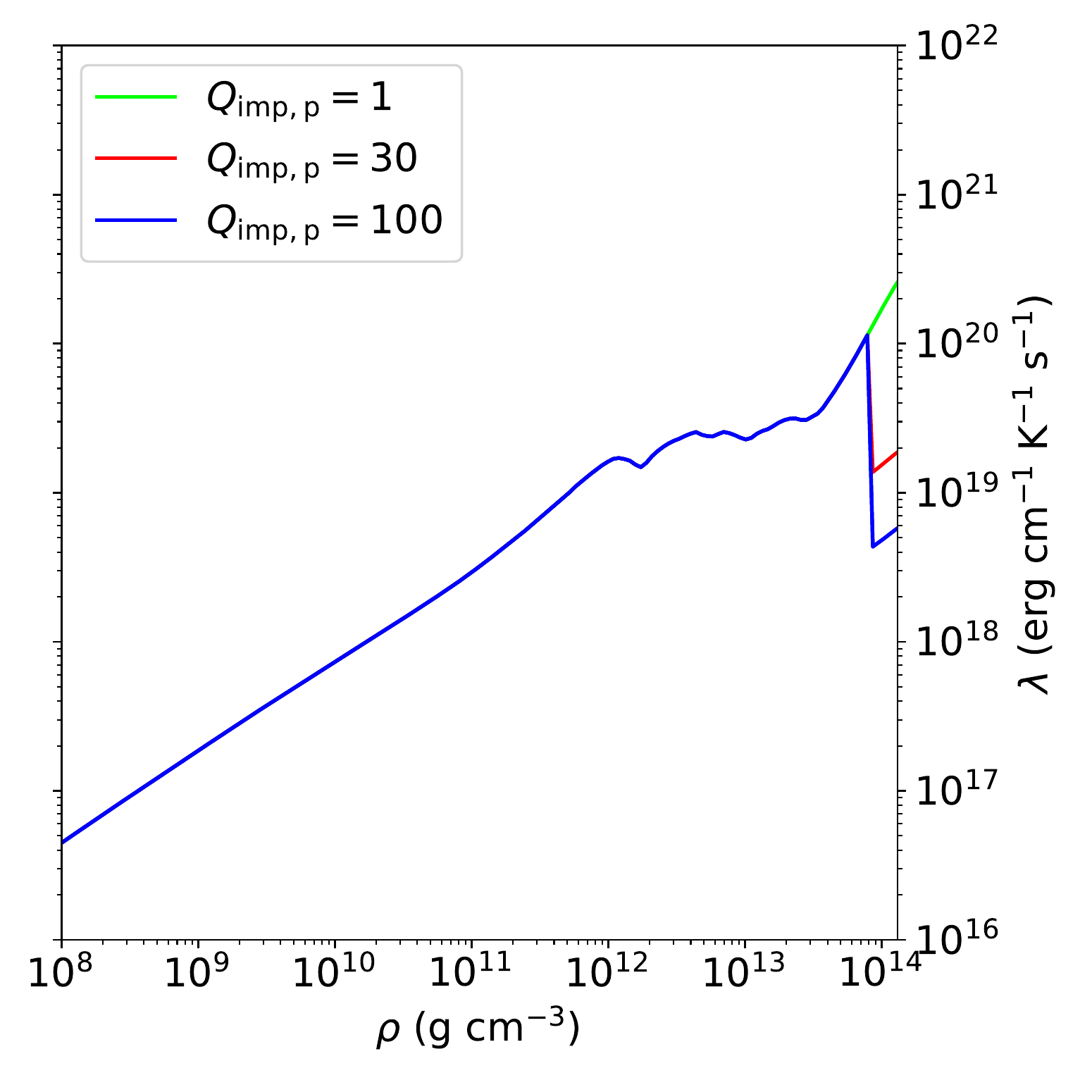}
     \caption{Thermal conductivity in the crust for different models. Left: Thermal conductivity in the crust due to electron-ion scattering (black dashed line), and electron-impurity scattering (dotted lines) for three different assumed impurity parameters throughout the crust, at a constant local temperature of $T=10^8$~K. At this temperature matter is crystallised in the whole density range displayed in this figure. The solid curves show the total thermal conductivity for the three models. For the green curves, an impurity factor of one is assumed in the crust (as in model 1). For the red curves, $Q_\text{imp}=30$ is assumed (as in model 16), and for the blue curves the impurity factor is set to $Q_\text{imp}=100$ (as in model 17). The contribution of electron-ion scattering to the total thermal conductivity is constant for the three models as indicated by the dashed black line. The total thermal conductivity consists of the contributions from electron-ion scattering, electron-impurity scattering, and electron-electron scattering, but the contribution of the last process in insignificant here. Right: Total thermal conductivity in the crust at a constant local temperature of $T=10^8$~K, for three models that assume different impurity factors for $\rho>8\times10^{13}\text{ g cm}^{-3}$, representing the pasta layers. For the green model, $Q_\text{imp,p}=1$ is assumed, for the red model, $Q_\text{imp,p}=30$ (as for model 18), and for the blue model, $Q_\text{imp,p}=100$ (as in model 19). In all models $Q_\text{imp}=1$ is assumed at lower densities in the crust. }
     \label{fig:lambda}
\end{figure*}

These results also have consequences for the position of observations from neutron star LMXBs on the $L_\text{q}-\langle\dot{M}\rangle$ plot, where $L_\text{q}$ is the observed quiescent luminosity \citep{yakovlev2004,heinke2009,heinke2010,wijnands2013,wijnands2017,han2017}. The heating curves with which the observations are compared to determine which cooling mechanisms might be at play, generally assume that ${L_\text{q}=\langle L_\text{H}\rangle -\langle L_\nu\rangle}$, with ${Q_\text{nuc}=1-2\text{ MeV nucleon}^{-1}}$ for the assumed heating rate. However, if the contribution from shallow heating is taken into account, the heating luminosity, and hence $L_\text{q}$ , is higher. It should then be possible to find sources that have a quiescent luminosity that exceeds the predicted photon cooling curve.

\subsection{Changes in the composition of the neutron star envelope}
In models 14 and 15 we varied the envelope composition compared to model 1, in which we assumed ${y_L=10^9\text{ g cm}^{-2}}$. In model 14 we decreased the light element layer to ${y_L=10^6\text{ g cm}^{-2}}$ and in model 15 we used a thicker layer of light elements (${y_L=10^{11}\text{ g cm}^{-2}}$). For each model we assumed that the light element layer consist of a helium layer on top of a carbon layer with ${y_\text{He}=0.1y_\text{C}}$. 

The results show that these variations do not cause significant changes in the evolution of the core temperature. For the three models shown in Fig.~\ref{fig:env} the difference between the models is most prominent in the surface temperature and photon luminosity. This is as expected, since the only difference between the models is how easily the heat that flows towards the surface (while most of the heat flows towards the core) can pass the envelope layer. For the model with the thickest layer of light elements, the thermal conductivity in the envelope is the highest, and therefore heat can leave the star most easily. Consequently, this star has a significantly higher effective temperature for similar crust temperatures. For models 14 and 15, there is only a very small difference in the equilibrium core temperature (a difference of $\sim1\%$ compared to model 1, see Table~\ref{tab:eqtemps}). This difference in equilibrium core temperature is caused by the fact that for the smaller light element layers less heat can flow away from the surface over an outburst cycle, so that relatively more heat flows towards the core. However, this effect is small, since for all envelope compositions most of the accretion-induced heat flows towards the core. %

\subsection{Assuming different thermal conductivity in the crust}
\begin{table*}
\caption{Neutron star properties for models 1, 12, and 13: stellar mass ($M$), stellar radius ($R$), crust thickness ($\Delta\,  R$), surface redshift factor ($e^{\phi_s}$), surface gravity ($g_\text{s}$), and central density ($\rho_\text{c}$).}
\label{tab:masses}
\centering\begin{tabular}{lcccccc}
\hline\hline
Model & $M$       & $R$  & $\Delta\,  R$ & $ e^{\phi_s}$ &  $g_\text{s}$        & $\rho_\text{c}$      \\
      & (M$_\odot$) & (km)   & (km)         &  & ($10^{14}$ cm s$^{-2}$)  & ($10^{15}$ g cm$^{-3}$) \\ \hline
12    & 1.2       & 11.8 & 1.4    & 0.84  & 1.4                 & 0.880                \\
1      & 1.6       & 11.5 & 0.9    & 0.77  & 2.1                 & 1.127                \\
13    & 2.0       & 11.0 & 0.5    & 0.68  & 3.2                 & 1.591    \\\hline           
\end{tabular}
\end{table*}

The next property that we consider is the thermal conductivity in the crust. By increasing the impurity factor of the crust, impurity scattering is enhanced, which decreases the thermal conductivity. In model 1 the impurity factor was set to a low value ($Q_\text{imp}=1$) throughout the crust. Typical values used for modelling crust cooling are on the order of $Q_\text{imp}\sim1$ up to $Q_\text{imp}\sim10$ \citep[e.g.][]{brown2009,merritt2016,turlione2015,cumming2017,parikh2019mxb}, although some sources may have higher impurity (regions) in the crust \citep{degenaar2014,ootes2019}. In models 16 and 17 the impurity was increased to $Q_\text{imp}=30$ and $Q_\text{imp}=100,$ respectively. This decreases the thermal conductivity in the crust, primarily for $\rho\gtrsim10^{11}\text{ g cm}^{-3}$, because the effect on the thermal conductivity at lower densities in the crust is largely dominated by electron-ion scattering (depending on the temperature and value of $Q_\text{imp}$, see left panel of Fig.~\ref{fig:lambda}). 

Figure~\ref{fig:Qimp} shows that the increased impurity (lower thermal conductivity) has a very small effect on the thermal evolution of the core. The effect is similar to what was determined for variation in the envelope composition and shallow heating depth. For lower thermal conductivity in the crust (higher $Q_\text{imp}$), the core equilibrium temperature is somewhat lower, but the difference in comparison to the basic model is small; $<1\%$ for model 16  and $<2\%$ for model 17 (see Table \ref{tab:eqtemps}). Although the thermal conductivity in the crust is decreased for higher $Q_\text{imp}$, the thermal conductivity in the inner crust is still high compared to the outer crust, and the direction of the heat flow is only slightly altered. Most of the heat generated in the crust is transported to the core in the crust-cooling phase. For the basic model, the internal luminosity at the crust-core boundary is 11 times higher than at the crust-envelope boundary once the long-term equilibrium is reached (at the time in each cycle of outbursts at which we measure the thermal properties, which is during the crust-cooling phase). Because the thermal conductivity is mostly decreased in the inner crust by increased $Q_\text{imp}$, heat flow towards the core is slower compared to the low-impurity crust, and more heat flows towards the surface (as can been seen from the photon luminosity curve in the right panel in Fig.~\ref{fig:Qimp}). However, even though the luminosity (once the long-term equilibrium is reached) at the crust-envelope boundary is almost twice as high for model 17 compared to model 1, the luminosity at the crust-core interface is still six times higher than at the crust-envelope boundary. All in all, despite the lower thermal conductivity assumed in models 16 and 17, the variation in the amount of heat that flows towards the core rather than the surface is small (in all cases most of the heat flows towards the core), causing the equilibrium core temperatures of these models to be similar to the model that assumes high thermal conductivity.

The right panel in Fig.~\ref{fig:Qimp} shows that at early times (${\lesssim200}$~yr), the neutrino luminosity is higher for the models with lower thermal conductivity in the crust (models 16 and 17). Due to the lower crustal thermal conductivity for models 16 and 17, the temperature in the crust at the time at which the model quantities are measured is higher than the crustal temperature in model 1 at that time (see left panel in Fig.~\ref{fig:Qimp}). This causes the neutrino emissivity in the crust to be high for models 16 and 17, and, moreover, dominant over the core neutrino processes, because the core temperatures at that time are still low. The difference disappears when the core of the neutron star becomes hot enough that the core neutrino emission dominates over that of the crust.

In models 18 and 19 we investigate if a low conductivity pasta layer (with respectively $Q_\text{imp,p}=30$ and $Q_\text{imp,p}=100$ at ${\rho>8\times10^{13}\text{ g cm}^{-3}}$) in the crust is able to affect the heat flow significantly. In these models the conductivity in the outer regions was still assumed to be high (${Q_\text{imp}=1}$). One might expect that a high-impurity pasta layer can provide a sufficient heat barrier to force less heat to flow to the core. However, Fig.~\ref{fig:Qimppasta} shows that this is not the case. The difference in equilibrium core temperature between models 1, 18, and 19 is negligible. The locally increased impurity does provide a heat barrier, as the thermal conductivity close to the crust-core boundary decreases by a factor of $\sim15$ and $\sim50$ for models 18 and 19, respectively, compared to model 1 (see right panel of Fig.~\ref{fig:lambda}). The heat flow in this specific region is thus slowed down, causing the crust to be hotter compared to model 1 in the crust-cooling phase (as can be observed from the surface temperature in Fig.~\ref{fig:Qimppasta}). However, the heat barriers in these two models are not strong enough to prevent heat from flowing into the core. Even when the impurity factor in the pasta layer is increased to $Q_\text{imp}=1000$, the difference in equilibrium core temperature compared to model 1 is only $\sim1\%$.

\subsection{Variation in stellar compactness}
In this subsection we will discuss the effect that stellar compactness has on the thermal evolution of transiently accreting neutron stars. However, we restrict ourselves here to cases that assume the minimal cooling paradigm \citep[i.e. no direct Urca emission, see][]{page2004}. Enhanced cooling will be discussed in Sect. \ref{sec:durca}. In our basic model we assumed a neutron star with mass ${M=1.6\text{ M}_\odot}$ and radius ${R=11.5\text{ km}}$. Such a neutron star has a surface gravity ${g_\text{s}=2.1\times10^{14}\text{ cm s}^{-2}}$, a crust thickness ${\Delta\,  R=0.9\text{ km}}$, a surface redshift $e^\phi_\text{s}=0.77$, and a central density ${\rho_\text{c}=1.127\times10^{15}\text{ g cm}^{-3}}$. In model 12 we assumed a lower mass of ${M=1.2\text{ M}_\odot}$ and in model 13 a heavy neutron star with a mass of ${M=2.0\text{ M}_\odot}$ was assumed. The stellar properties of these models are summarised in Table~\ref{tab:masses}.

\begin{figure*}
 \includegraphics[width=\textwidth]{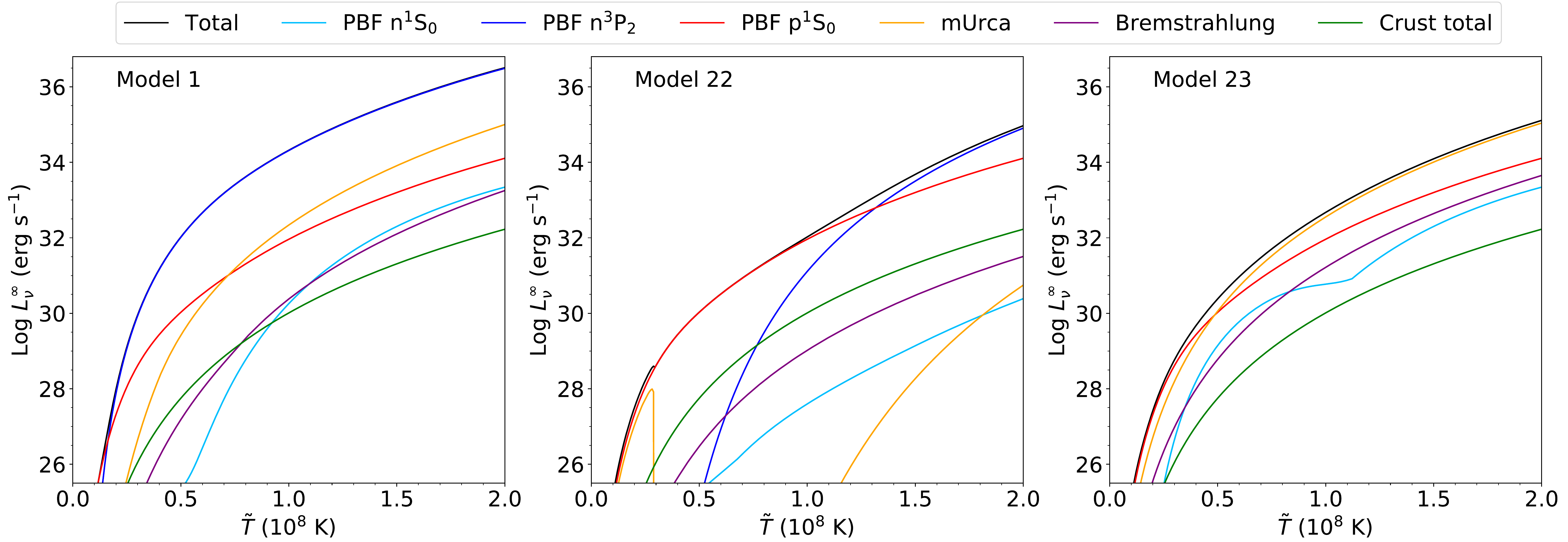}
    \caption{Neutrino luminosity of different processes as a function of temperature as used in {\tt NSCool}, for three models assuming different neutron $^3\text{P}_2$ pairing gaps. In model 1 gap model `a' is assumed (left panel), in model 22 gap model `c' is assumed (middle panel), and in the model 23 neutron $^3\text{P}_2$ pairing is omitted (right panel). }
    \label{fig:Lnu_n3p2}
\end{figure*}

The results presented in Fig.~\ref{fig:masses} show that for smaller masses (smaller compactness), the core reaches higher equilibrium temperatures and that the equilibrium state is reached faster than for heavier stars (${\tilde{T}_\text{core,eq}=1.275\times10^8\text{ K}}$ and ${t_\text{eq}=4.3\times10^3}$~yr for model 12, and ${\tilde{T}_\text{core,eq}=1.007\times10^8\text{ K}}$ and $t_\text{eq}=16.2\times10^3$~yr for model 13, see Table \ref{tab:eqtemps}). For a higher stellar mass, the neutron star has a higher central density. For the chosen neutron $^3\text{P}_2$ pairing gap in our models, this means that the critical temperature curve for the pairing gap is cut off in the centre at temperatures $T_\text{crit}\sim 3\times 10^8$~K for the $1.2\text{ M}_\odot$ star, while the curve continues to lower temperatures for the heavier stars ($T_\text{crit}(\rho_\text{c})=1.3\times10^8$~K for $M=1.6\text{ M}_\odot$ and $T_\text{crit}(\rho_\text{c})=0.3\times10^8$~K for $M=2.0\text{ M}_\odot$). This has consequences for both the specific heat and neutrino luminosity of these stars. On the one hand, the specific heat as a function of temperature (over the range of temperatures considered here), is higher for the higher mass stars, because neutron pairing enhances the specific heat for the range $T/T_\text{crit}\sim[0.5-1]$ \citep[see][]{page2004}. As a consequence, the amount of heat needed to heat up the core is larger, and therefore the equilibrium state is reached over a longer timescale. On the other hand, the neutrino luminosity is stronger for higher mass stars at the same temperature (as can be determined from comparison of the left and right panels in Fig.~\ref{fig:masses}), because PBF neutrino emission becomes significant (and consequently dominant over all other neutrino emission processes in these models) for core temperatures in the range $T/T_\text{crit}\sim[0.2-1.0]$. This causes an equilibrium to be established at lower core temperatures for the models in which a higher mass is assumed. Finally, the difference in neutrino luminosity as a function of temperature for the different models, in combination with the different specific heat as a function of temperature, is what causes the neutrino luminosity curves in the right panel of Fig. \ref{fig:masses} to intersect.

\subsection{Assuming different pairing characteristics}

In this subsection we focus on pairing characteristics and their effects on the neutrino emission. In model 1 for the pairing of dripped neutrons in the crust, we used the `SFB' gap model from \citet{schwenk2003}. In model 20 we used instead the neutron $^1\text{S}_0$ gap model `CCDK' calculated by \citet{chen1993}, which has a maximum critical temperature $T_\text{crit,max}\sim6\times10^9\text{ K}$. Contrary to the gap model used in model 1, the one assumed in model 20 closes inside the crust at a density $\sim9\times10^{13}\text{ g cm}^{-3}$. For a neutron $^1\text{S}_0$ gap that closes in the crust, a layer of non-paired neutrons may exist at the bottom of the crust as long as the neutron $^3\text{P}_2$ gap starts at a higher density. Such a layer affects the heat flow towards the core \citep[see the discussion by][]{deibel2017}. In model 20 we force the existence of a layer of non-paired neutrons at the bottom of the crust by cutting off the $^3\text{P}_2$ pairing gap at the crust-core boundary. In model 21 we assumed that the crust is completely non-superfluid. The effects of the two models compared to the basic model are very small; the equilibrium core temperature of models 20 and 21 is $\sim0.1\%$ lower than for model 1. A small difference in core temperature evolution over the period in which the equilibrium state is reached can be observed from Fig.~\ref{fig:n1s0}. Due to the pairing of neutrons in the inner crust, the specific heat is lower in this region for models 1 and 20 than for model 21, such that the thermal diffusivity in this region is higher. The heat flow towards the core is somewhat higher and hence the core can heat up faster. This explains why for the models in which the free neutrons in the crust are paired, the equilibrium temperature is reached faster (see Table \ref{tab:eqtemps}; the equilibrium timescales for models 1 and 20 are respectively 4\% and 8\% lower compared to model 21). 

Next, we consider neutron $^3\text{P}_2$ pairing in the core of neutron stars. In model 1, we assumed the gap model `a' from \citet{page2004}, which has a maximum critical temperature ${T_\text{crit,max}=10^9}$~K. In model 22 we assumed the neutron $^3\text{P}_2$ gap model `c' from \citet{page2004} instead. This gap model has a higher maximum critical temperature of $T_\text{crit,max}=10^{10}$~K, such that the neutrons form stronger pairs for the typical temperatures reached in the core for these models. To consider the other extreme situation, we assumed in model 23 that no neutron $^3\text{P}_2$ pairing takes place. In that case the neutrons in the core are unpaired (but the protons can still form pairs), except for the outermost region of the core up to where the assumed neutron $^1\text{S}_0$ gap reaches. The neutron $^1\text{S}_0$ gap closes at a density ${\rho\sim1.8\times10^{14}\text{ g cm}^{-3}}$. For both models 22 and 23 the core reaches an equilibrium core temperature that is much higher than for our basic model ($\tilde{T}_\text{core,eq}=1.837\times10^8$~K for model 22 and $\tilde{T}_\text{core,eq}=1.732\times10^8$~K for model 23, see Fig.~\ref{fig:n3p2}). There are two factors that are responsible for the observed effects between these three models. First of all, there is a difference in specific heat between the three models. In models 1 and 22, the core specific heat is suppressed for $\tilde{T}<10^8$~K due to neutron pairing. Because the critical temperatures for the pairing gap assumed in model 22 are higher than for model 1, the specific heat suppression is stronger. As a consequence, for model 22 the equilibrium timescale is the shortest, and, vice versa, model 23 (no pairing, and hence the highest specific heat) has the lowest core heating rate. Secondly, there are large differences in neutrino emission from the PBF process between these models (see Fig.~\ref{fig:Lnu_n3p2}). This neutrino emission process is -- in the absence of direct Urca emission -- the strongest neutrino emission process, and it becomes important when the temperature increases and reaches values close to the critical temperature for nucleon pairing \citep[see e.g. Fig.~5 in ][for the control function for this process as a function of $T_\text{crit}$]{page2006}. For gap `a' \citep[][assumed in model 1]{page2006}, the neutron pairing gap comprises the entire core with critical temperatures in the range $(1.4\times10^8-1.0\times10^9)$~K, while for gap `c' \citep[][assumed in model 22]{page2006} the critical temperatures are in the range ($7.2\times10^8-1.0\times10^{10}$~K). This means that in model 1 the PBF process becomes efficient at much lower core temperatures than for model 22, such that equilibrium between heating and cooling is reached at much lower core temperatures (see Fig.~\ref{fig:Lnu_n3p2}). For model 23, neutrino emission from the PFB process for neutron $^3\text{P}_2$ is absent, and it is the modified Urca process that controls the equilibrium core temperature. Because the neutrons in the core are unpaired, the modified Urca luminosity is higher as a function of temperature than for models 1 and and 22. The total neutrino luminosity as a function of temperature is higher in model 23 than in model 22, and hence equilibrium is reached at a lower core temperature.

The last pairing gap to consider is the proton $^1\text{S}_0$ gap, which controls the pairing of the small fraction of protons in the core of neutron stars. We compare here again the pairing gap assumed in model 1 with a model in which proton pairing is stronger (model 24) and a model in which protons are assumed to be unpaired (model 25). In model 1 we assumed the proton pairing gap from \citet{amundsen1985}, which has a critical temperature function that reaches from the crust-core boundary up to a density $\rho=9.7\times10^{14}\text{ g cm}^{-3}$. The function peaks at low density in the core and has a maximum critical temperature $T_\text{crit,max}=2.3\times10^9$~K. The gap model assumed in model 24 is from \citet{chen1993}. It comprises the full core and has a maximum critical temperature of $T_\text{crit,max}=6.6\times10^9$~K. The results in Fig.~\ref{fig:p1s0} show that the assumptions for proton pairing do not affect the equilibrium core temperature. As can be seen from the left panel in Fig.~\ref{fig:Lnu_n3p2}, the neutrino emission from the proton PBF process is negligible compared to the neutron PBF emission in model 1 at temperatures $\tilde{T}\sim10^8$~K. This depends on the assumed gap model and on the temperature. For model 24 the contribution from proton PBF neutrino emission at $\tilde{T}\sim10^8$~K to the total neutrino emissivity is similar compared to model 1 and absent for model 25, such that for both models the total neutrino emission at the temperature at which equilibrium is reached is the same as for model 1 (see right panel in Fig.~\ref{fig:p1s0}). However, there is a difference in the timescale over which equilibrium is reached between models 24, 25, and the basic model due to differences in specific heat. At temperatures $\tilde{T}\lesssim10^8$~K, the specific heat for the two models that assume proton pairing (model 1 and model 24) is reduced because the internal temperature is significantly lower than the critical temperature for the assumed proton gap models. Because the critical temperatures assumed in model 24 are the highest, the specific heat reduction in model 24 is stronger than for model 1. The neutron star in model 24 can therefore heat up most easily, as can be observed from Fig.~\ref{fig:p1s0}. Similarly, because no proton pairing takes place in model 25, this model has the highest specific heat at low temperatures
such that the star heats up more slowly and the equilibrium timescale is higher than for models 1 and 24. 

 \begin{figure}
  \includegraphics[width=0.5\textwidth]{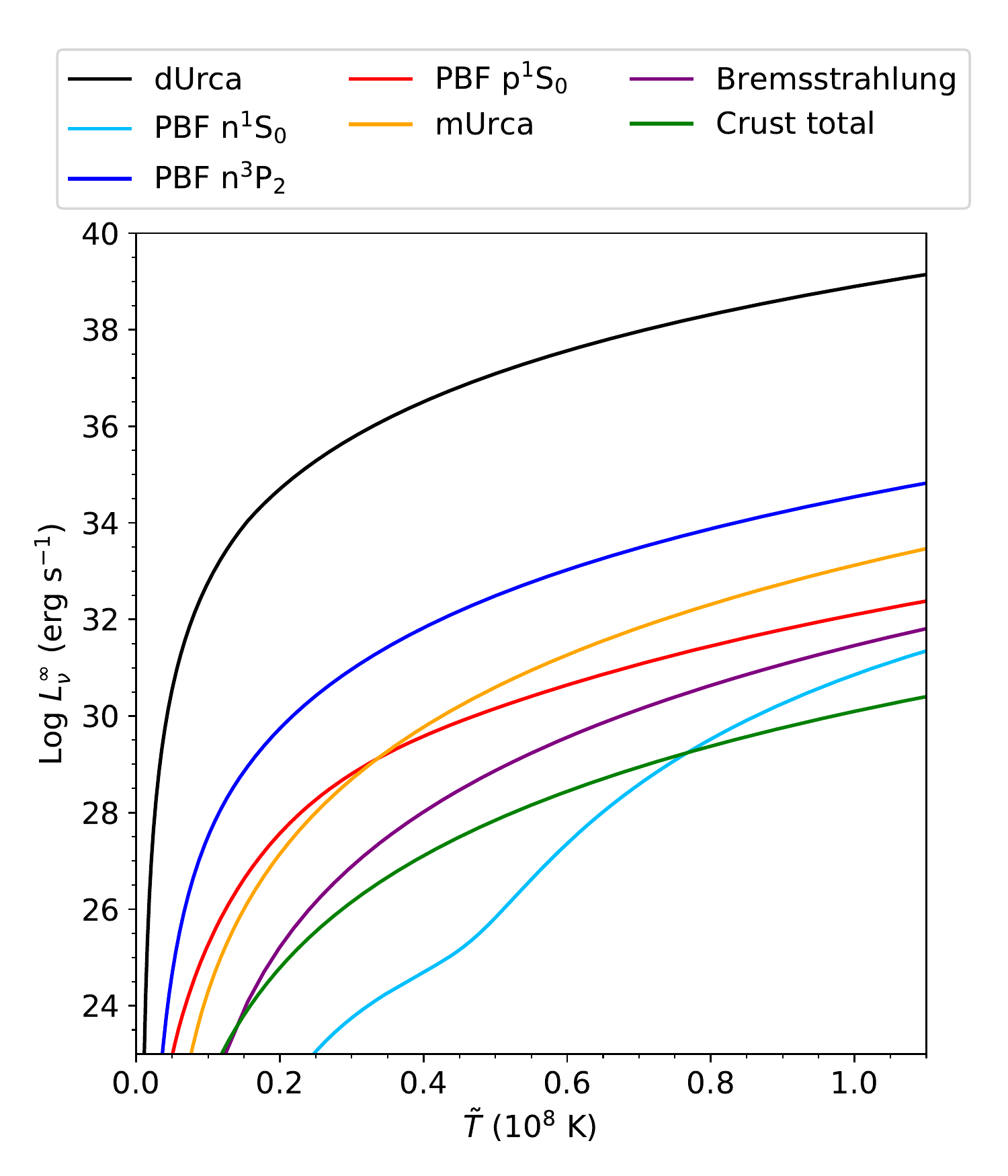}
     \caption{Neutrino luminosities as a function of temperature used in {\tt NSCool} for model 26.}
     \label{fig:lumdurca}
\end{figure}

\subsection{Direct Urca emission}\label{sec:durca}

The direct Urca process is the most efficient neutrino emission process that might take place in the cores of heavy neutron stars, under the requirement that the fraction of protons is large enough \citep[$\sim11\%$;][]{lattimer1991}. In model 26 we assumed a neutron star with a mass $M=2.0\text{ M}_\odot$, which allows for direct Urca to take place in the innermost part of the core at densities $\rho\geq1.257\times10^{15}\text{ g cm}^{-3}$. The direct Urca emission region covers $\sim1$\% of the total volume of the star. 

In Fig.~\ref{fig:durca} we compare the results of this model with model 13, in which a neutron star with the same mass is assumed, but where the direct Urca emission process is not taken into account. In model 26 the neutron star reaches a very low equilibrium core temperature of $\tilde{T}_\text{core,eq}=0.194\times10^8$~K in $t_\text{eq}=300$~yr. 
Figure~\ref{fig:lumdurca} shows the neutrino emission luminosity as a function of temperature as used in {\tt NSCool} for model 26, which shows that direct Urca emission is orders of magnitude larger than any of the other neutrino emission processes. Due to the fast increase in luminosity with temperature, the neutrino emission from direct Urca emission is at low temperatures already strong enough to balance the heating rate from the accretion outbursts. The equilibrium timescale of model 26 is more than 50 times lower than that of model 13. If direct Urca takes place in only a small fraction of the core such as might be the case for MXB 1659-29 \citep{brown2018}, the neutron star is very likely to be in equilibrium state.

\section{Conclusion}

In this parameter study we investigated the long-term temperature evolution of transiently accreting neutron stars. Assuming that such neutron stars have regular accretion outbursts, they will reach an equilibrium state in which the heat generated in the crust by accretion-induced reactions is balanced by neutrino cooling from the core and photon emission from the surface. We compared how the properties of neutron star crusts and cores, as well as their outburst properties affect the equilibrium state: the equilibrium core temperature and the time over which this temperature is reached. We used the observed outburst behaviour of Aql~X-1 over the past $\sim20$~yr and extrapolated the sequence of 24 outbursts observed in this period to model such outbursts for $10^5$~yr using our thermal evolution code {\tt{NSCool}}, assuming that the neutron star heats up from a cold state. The results are as follows.
\begin{itemize}
\item As expected, we find that outburst properties, such as the recurrence time and outburst accretion rate strongly affect the equilibrium state. These properties set the time-averaged accretion rate and thus affect the amount of heat that is injected into the neutron star over an outburst cycle. We find that the equilibrium core temperature as a function of equilibrium timescale follows a logarithmic decay trend, for decreasing accretion rate.
\item We find that shallow heating can significantly increase the equilibrium core temperature, independently of the depth of this heat source. Varying the depth of the shallow heat source has a non-significant effect on the equilibrium parameters. Even when the shallow heating is injected just below the envelope on the neutron star, most of the heat flows towards the core rather than the surface. The strength of the shallow heating therefore affects the long-term equilibrium state of neutron stars. When shallow heating is the only assumed heating process, the core will reach an equilibrium core temperature similar to the equilibrium core temperature for a model that only takes into account deep crustal heating (the difference is $2\%$ in our models if the same heating strength is assumed for the heat sources). From these models we conclude that it is possible to heat accreting neutron stars only by the shallow heating process. Additionally, stars with strong shallow heating may be observed to have higher quiescent luminosities than the calculated photon cooling curves, because these curves do not take into account the additional contribution of shallow heating to the heating rate of the star.
\item Under the assumption of minimal cooling and for the assumed pairing gaps in our basic model, increasing the stellar compactness increases the neutrino luminosity. Naturally, the stronger cooling by neutrino emission from the core in these models leads to a lower equilibrium core temperature for heavier (more compact) neutron stars. 
\item The envelope composition has a strong effect on the observed surface temperature neutron stars, but changing the thermal conductivity of this heat blanket (by changing the composition of the envelope) has marginal effects on the equilibrium state (differences in equilibrium core temperature are $\sim1\%$ compared to the basic model), which are not considered to be significant. 
\item Decreasing the thermal conductivity (by increasing the impurity factor) in the crust delays the heat flow, but even an impurity factor of $Q_\text{imp}=100$ is not strong enough to change the equilibrium state significantly. The difference in equilibrium core temperature is $<2\%$ compared to the model in which $Q_\text{imp}=1$ is assumed. Moreover, a low-conductivity pasta layer (with $Q_\text{imp}=30$ or $Q_\text{imp}=100$) cannot provide a heat barrier that prevents heat from flowing towards the core.
\item It is well known that the exact details of the assumed pairing gaps for neutron superfluidity and proton superconductivity in the core of neutron stars have strong consequences for the properties of the core. At $T\lesssim T_\text{crit}$ the pairing of nuclei affects the specific heat and the neutrino emission. However, both quantities are very sensitive to the local $T/T_\text{crit}$, and hence the effects on the equilibrium core temperature and the timescale over which equilibrium is reached depend strongly on the exact details of the assumed models for neutron and proton pairing. The pairing of dripped neutrons in the crusts of neutron stars has a minor effect on the equilibrium state. 

\item The strongest effect on the equilibrium state of neutron stars is provided by the direct Urca neutrino emission process. If the nucleons in the innermost region of the core of massive neutron stars are unpaired, and the proton fraction of the material is large enough, this process can take place. Because it has a much higher emissivity at a given temperature than any of the other neutrino emission processes, it can balance the heat that enters the core at much lower temperatures than when the minimal cooling processes are assumed to take place. Hence a neutron star in which direct Urca emission takes place will be much colder when it reaches the equilibrium state. We have shown here that in the presence of a direct Urca process the time to reach the equilibrium state is much shorter, by one to three orders of magnitude, and can be as small as a few hundred years, confirming the estimates of \protect\citet{wijnands2013}.
\end{itemize}

\begin{acknowledgements} 
LSO and RW received support from a NWO TOP Grant, module 1, awarded to RW.  
DP is partially supported by the Mexican Conacyt
(CB-2014-01, \#240512). 
This work was supported in part by the National Science Foundation under Grant No. PHY-1430152 (JINA Center for the Evolution of the Elements).
\end{acknowledgements}

\bibliographystyle{aa}
\bibliography{Core_T_evolution}

\appendix

 \begin{figure*}
 \section{Additional figures}
  \includegraphics[width=0.5\textwidth]{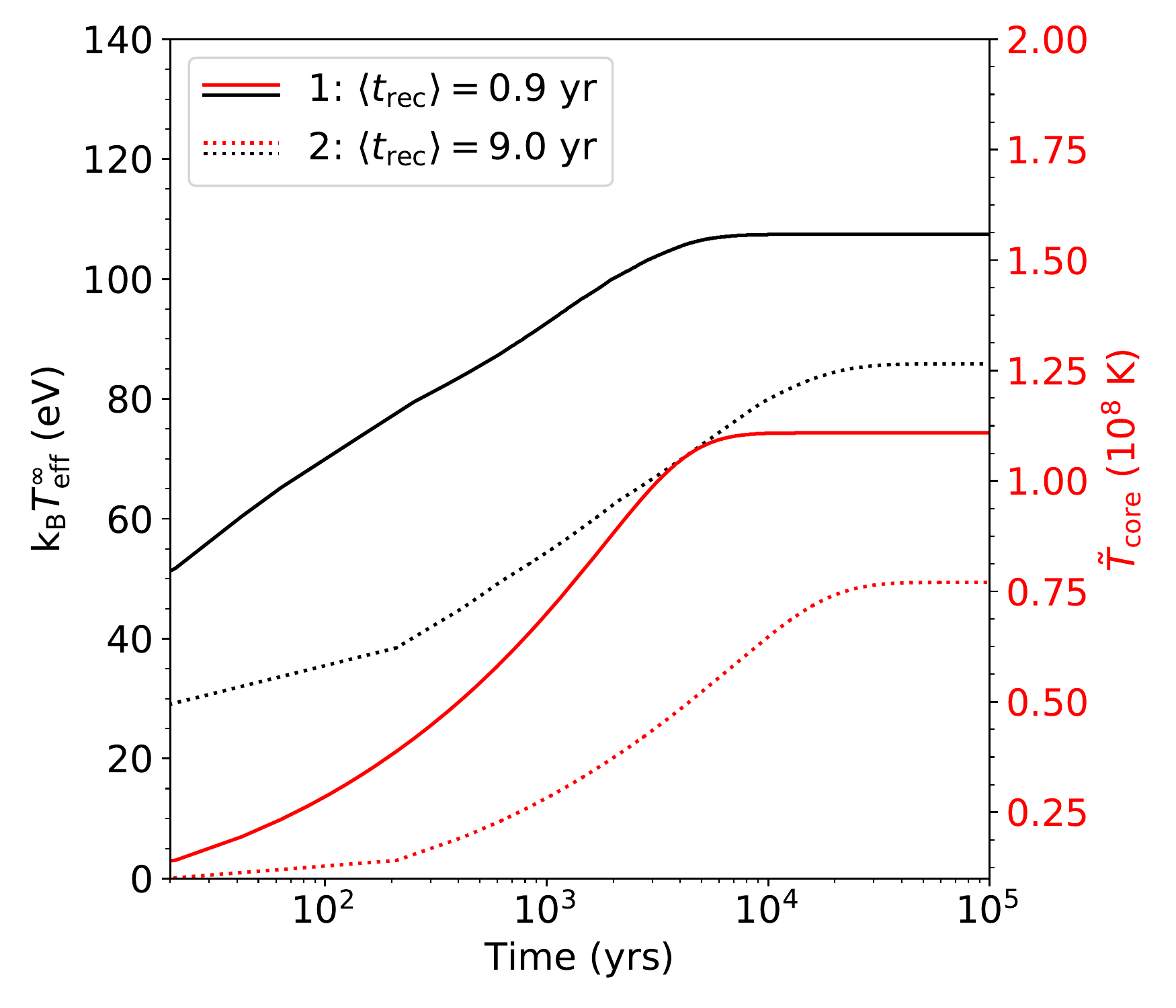}\includegraphics[width=0.5\textwidth]{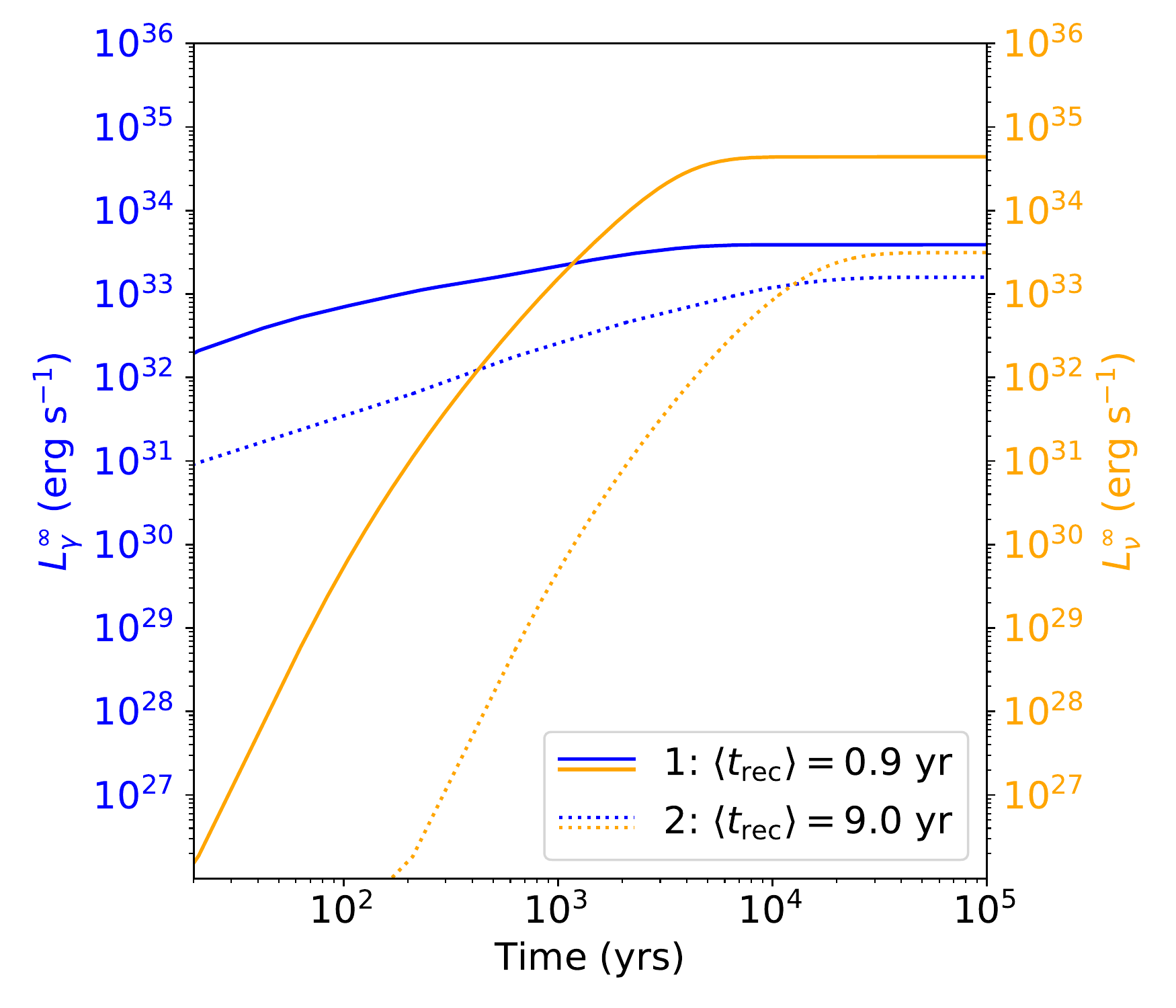}
     \caption{Results of models 1 and 2. Left: Evolution of the effective temperature (black curves) and core temperature (red curves) for model 1 (solid curves; our basic model) and model 2 (dotted curves). In the basic model (model 1), the outbursts have an average recurrence time of 0.9~yr, while in model 2 this is 9.0~yr. All other parameters of the two models are the same (including the duration of outbursts). Right: Evolution of the photon luminosity (blue) and neutrino luminosity (orange) in model 1 (solid curves) and model 2 (dotted curves).}
     \label{fig:trec}
\end{figure*}
 \begin{figure*}
  \includegraphics[width=0.5\textwidth]{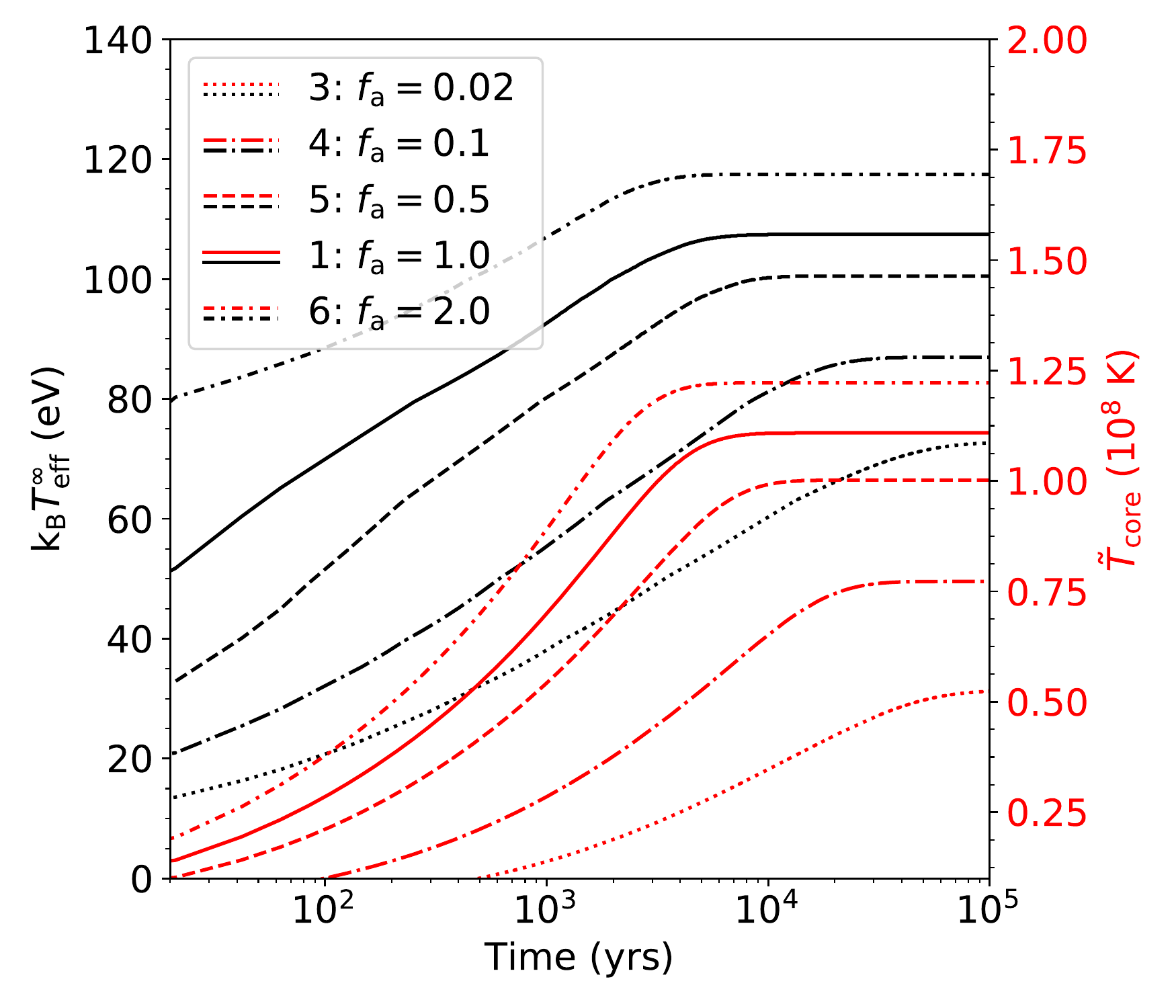}\includegraphics[width=0.5\textwidth]{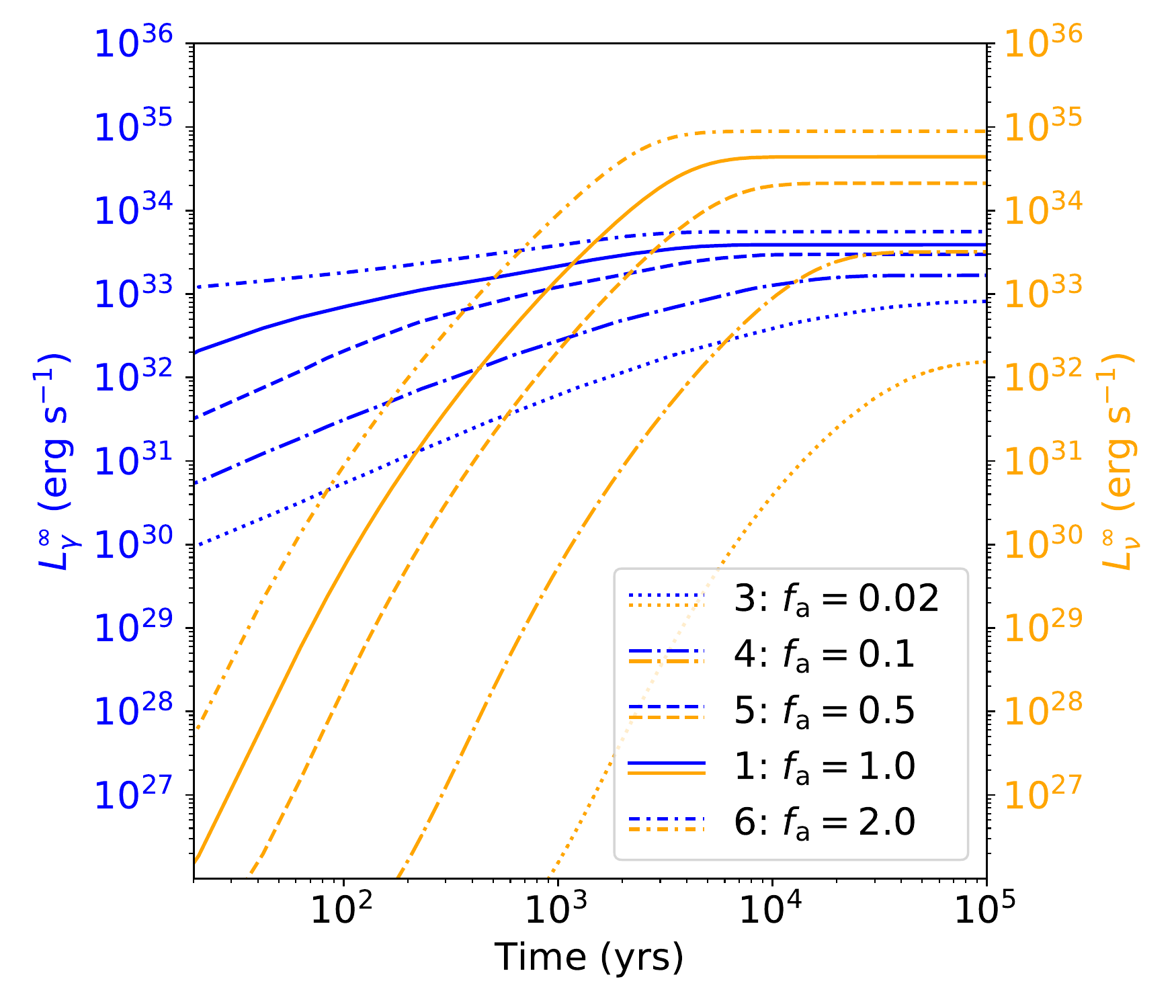}
     \caption{Same as Fig.~\ref{fig:trec}, but for variation in average outburst accretion rate. For the models shows in this figure, the outburst accretion rate ($\dot{M}_\text{o}$), is multiplied by a factor of $f_\text{a}$. The solid lines are for model 1 (which has a time-averaged accretion rate $\langle\dot{M}\rangle=3.2\times10^{-10}\text{ M}_\odot\text{ yr}^{-1}$), the dotted line for model 3 ($\langle\dot{M}\rangle=6.4\times10^{-12}\text{ M}_\odot\text{ yr}^{-1}$), the dashed-dotted line for model 4 ($\langle\dot{M}\rangle=3.2\times10^{-11}\text{ M}_\odot\text{ yr}^{-1}$), the dashed lines for model 5 ($\langle\dot{M}\rangle=1.6\times10^{-10}\text{ M}_\odot\text{ yr}^{-1}$), and the short dashed-dotted curves for model 6 ($\langle\dot{M}\rangle=6.4\times10^{-10}\text{ M}_\odot\text{ yr}^{-1}$).}
     \label{fig:mdot}
\end{figure*}

 \begin{figure*}
  \includegraphics[width=0.5\textwidth]{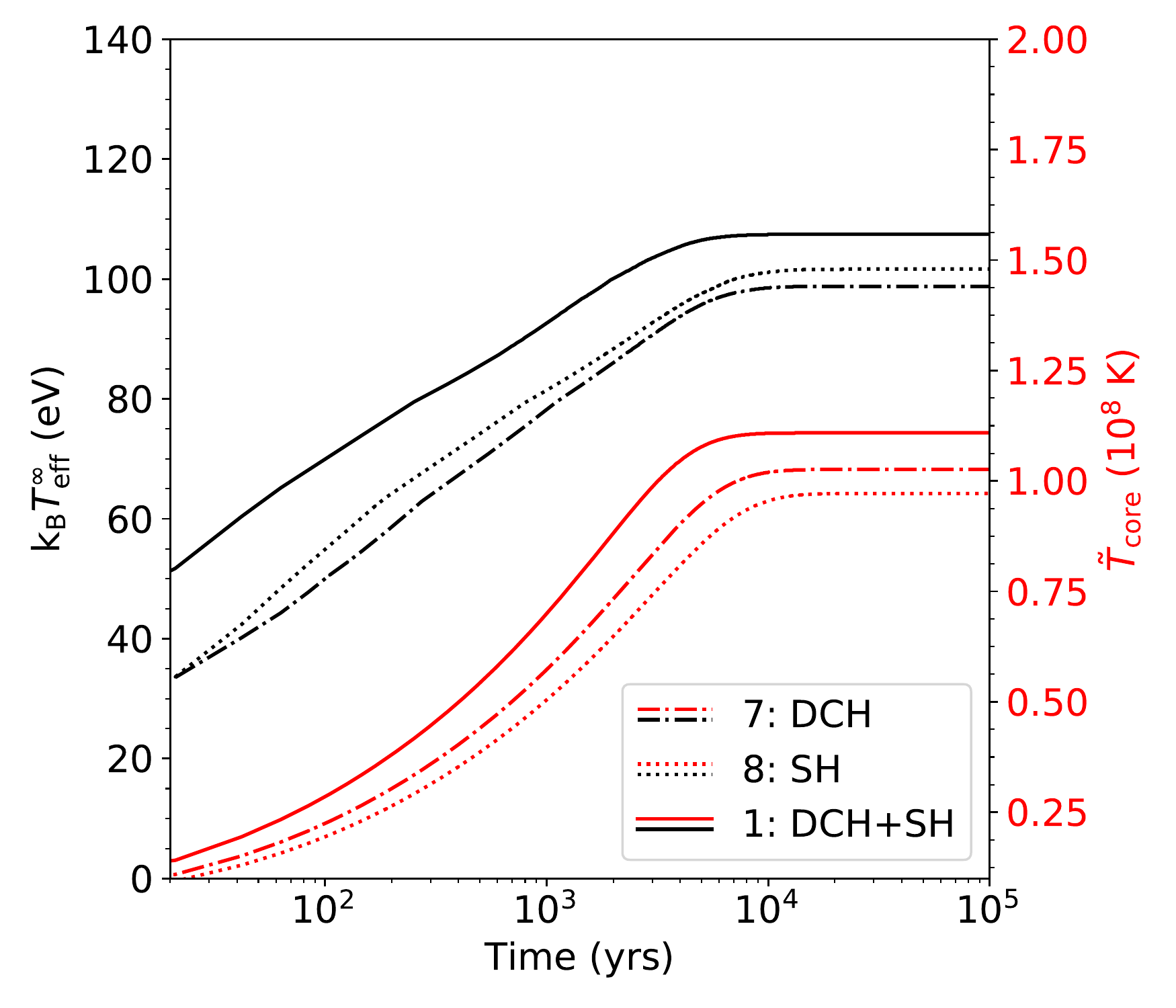}\includegraphics[width=0.5\textwidth]{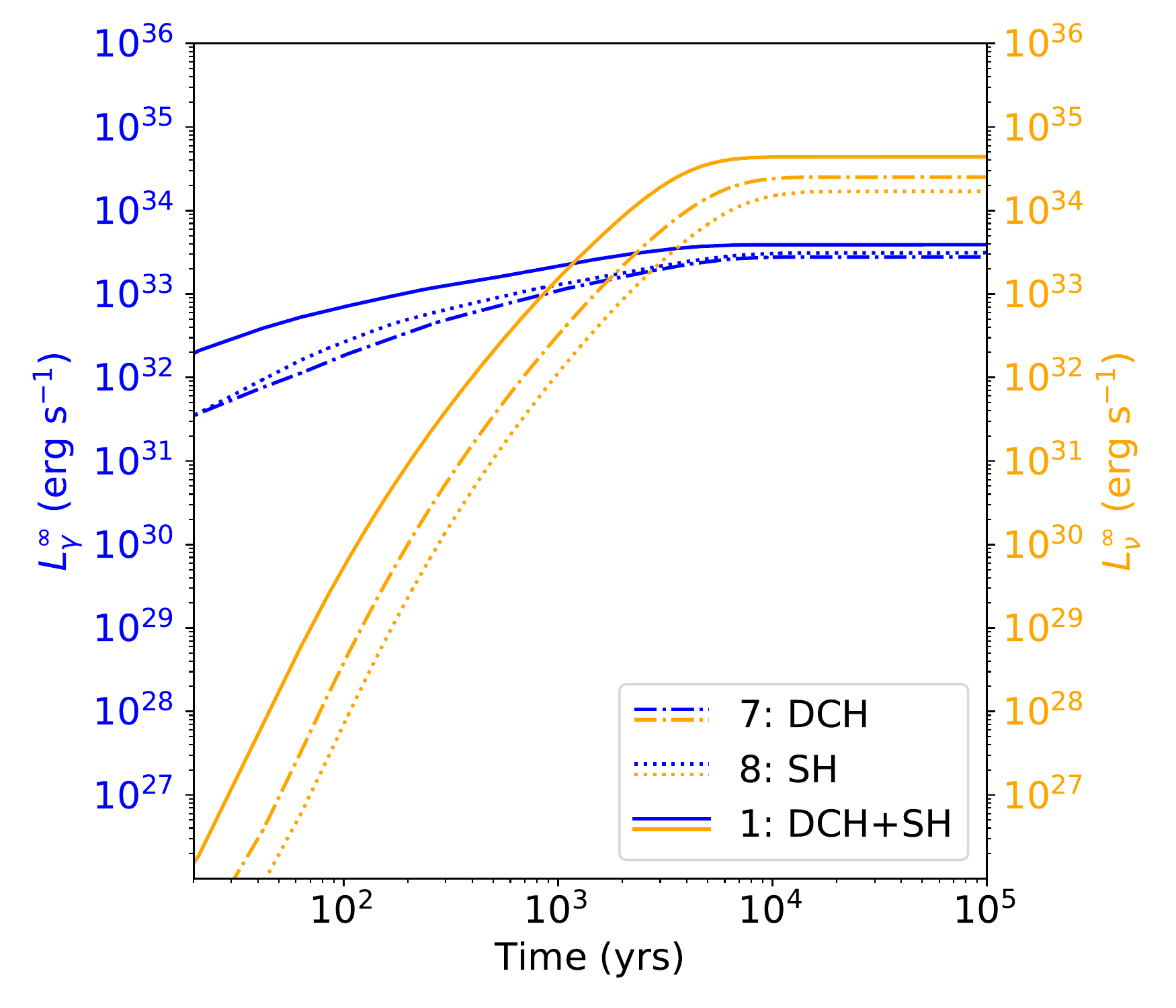}
     \caption{Same as Fig.~\ref{fig:trec}, but comparing the individual contribution of the shallow heating and deep crustal heating to that of the basic model. The solid line shows model 1, in which both deep crustal heating and shallow heating are assumed to be active. The dashed-dotted curves show the results of model 7, in which only deep crustal heating is assumed to be active (and shallow heating is inactive), and the dotted curves represent model 8, in which deep crustal heating is deactivated and only the shallow heating process takes place.}
     \label{fig:heating}
\end{figure*}

 \begin{figure*}
  \includegraphics[width=0.5\textwidth]{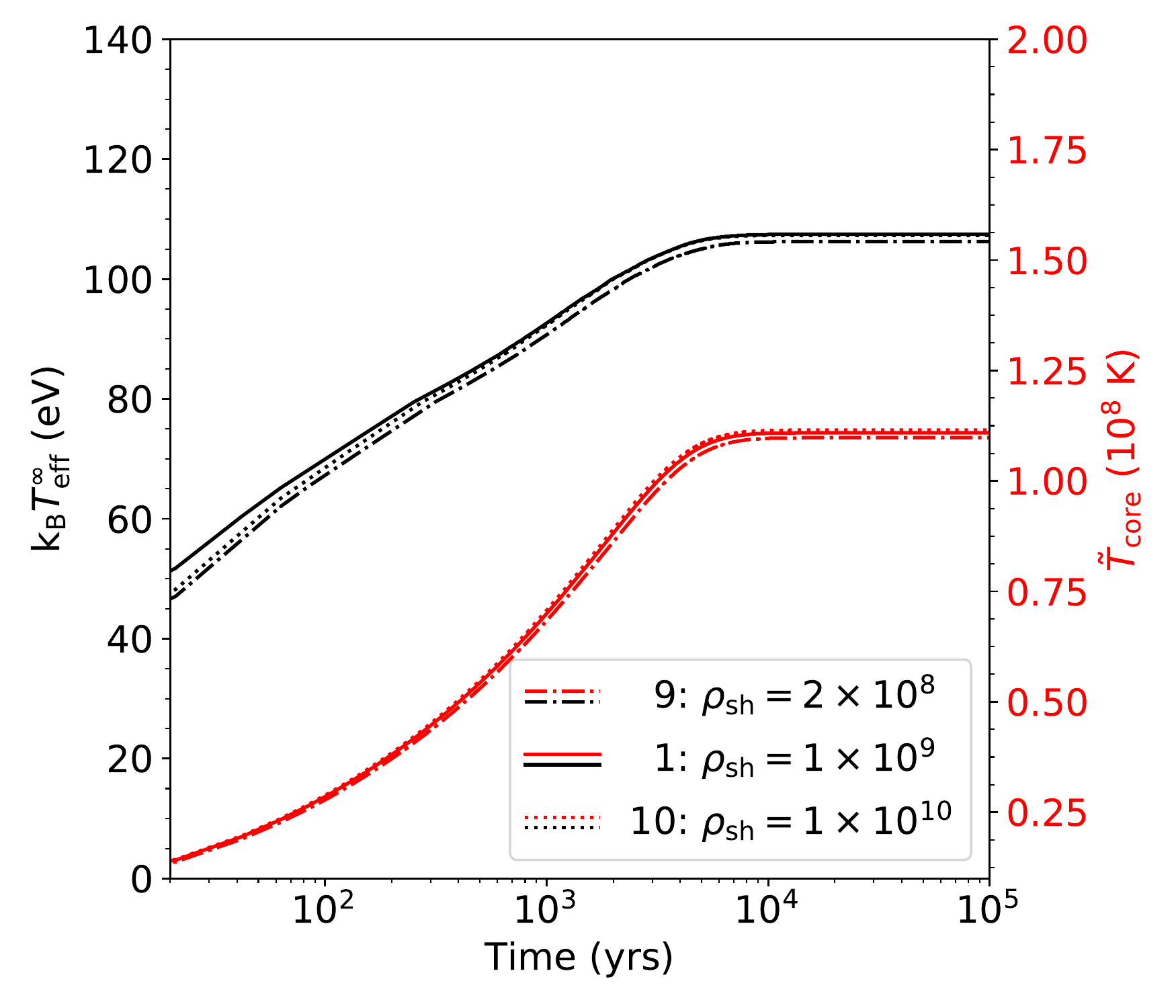}\includegraphics[width=0.5\textwidth]{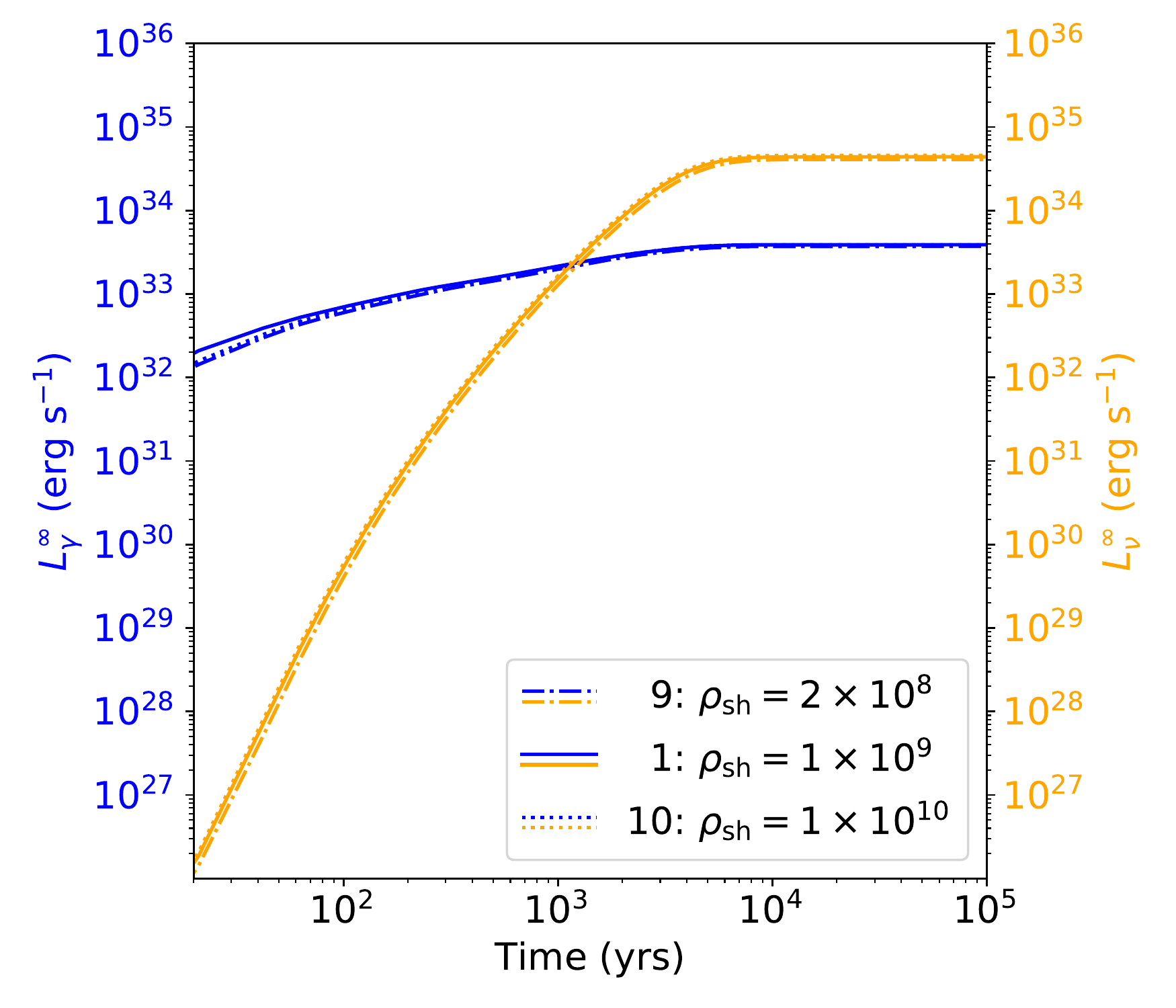}
     \caption{Same as Fig.~\ref{fig:trec}, but for models in which the shallow heating occurs at different depths in the outer crust. The solid line shows the results of model 1 (shallow heating in the density range ${\Delta\,  \rho_\text{sh}=[1-5]\times10^9\text{ g cm}^{-3}}$). The dashed-dotted curves shows the results for model 9, in which the shallow heating is placed closer to the surface, at a density range ${\Delta\,  \rho_\text{sh}=[2-10]\times10^8\text{ g cm}^{-3}}$, and the dotted curves show the results for model 10, in which the shallow heating is assumed to be located deeper in the crust (in the range ${\Delta\,  \rho_\text{sh}=[1-5]\times10^{10}\text{ g cm}^{-3}}$).
     }
     \label{fig:rhosh}
\end{figure*}

 \begin{figure*}
  \includegraphics[width=0.5\textwidth]{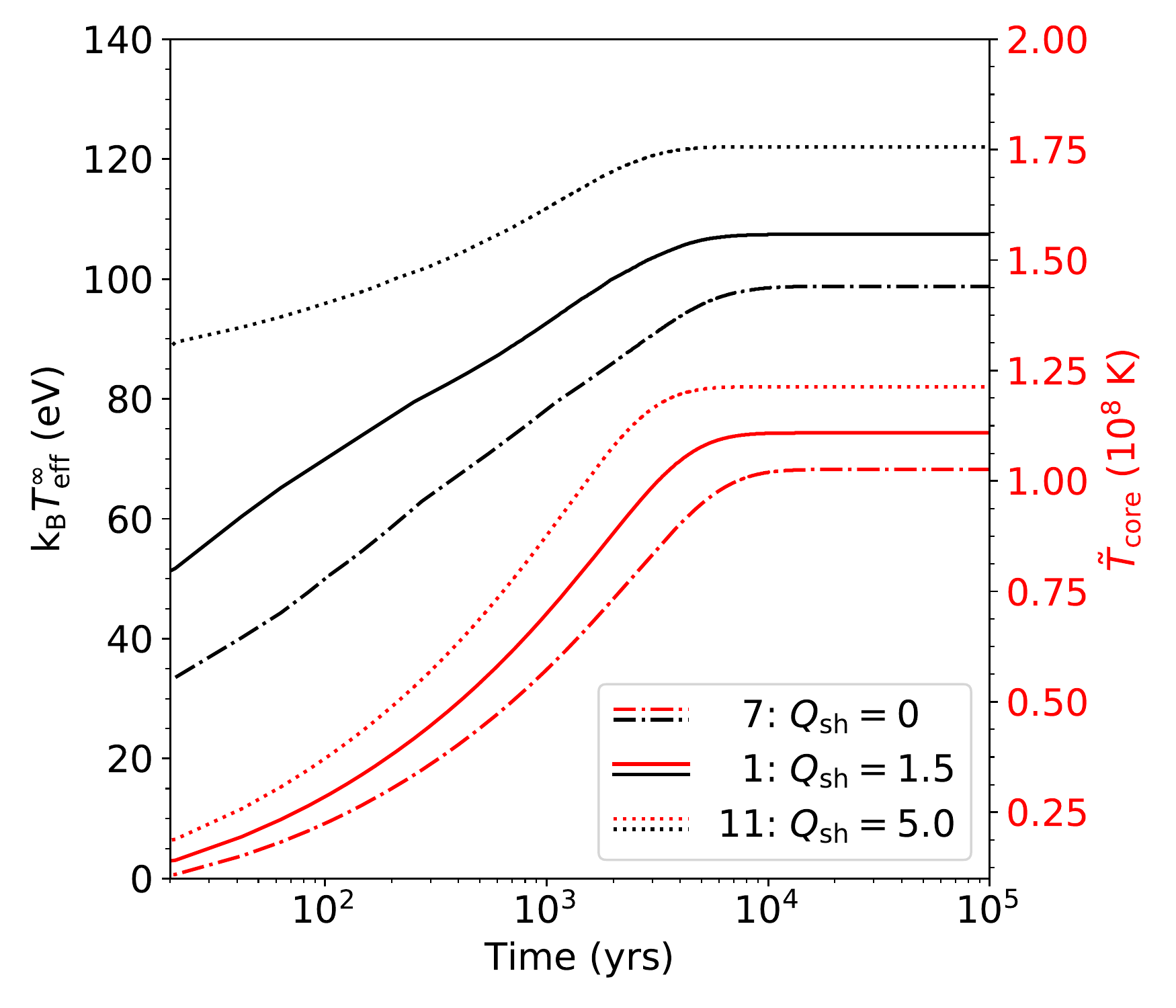}\includegraphics[width=0.5\textwidth]{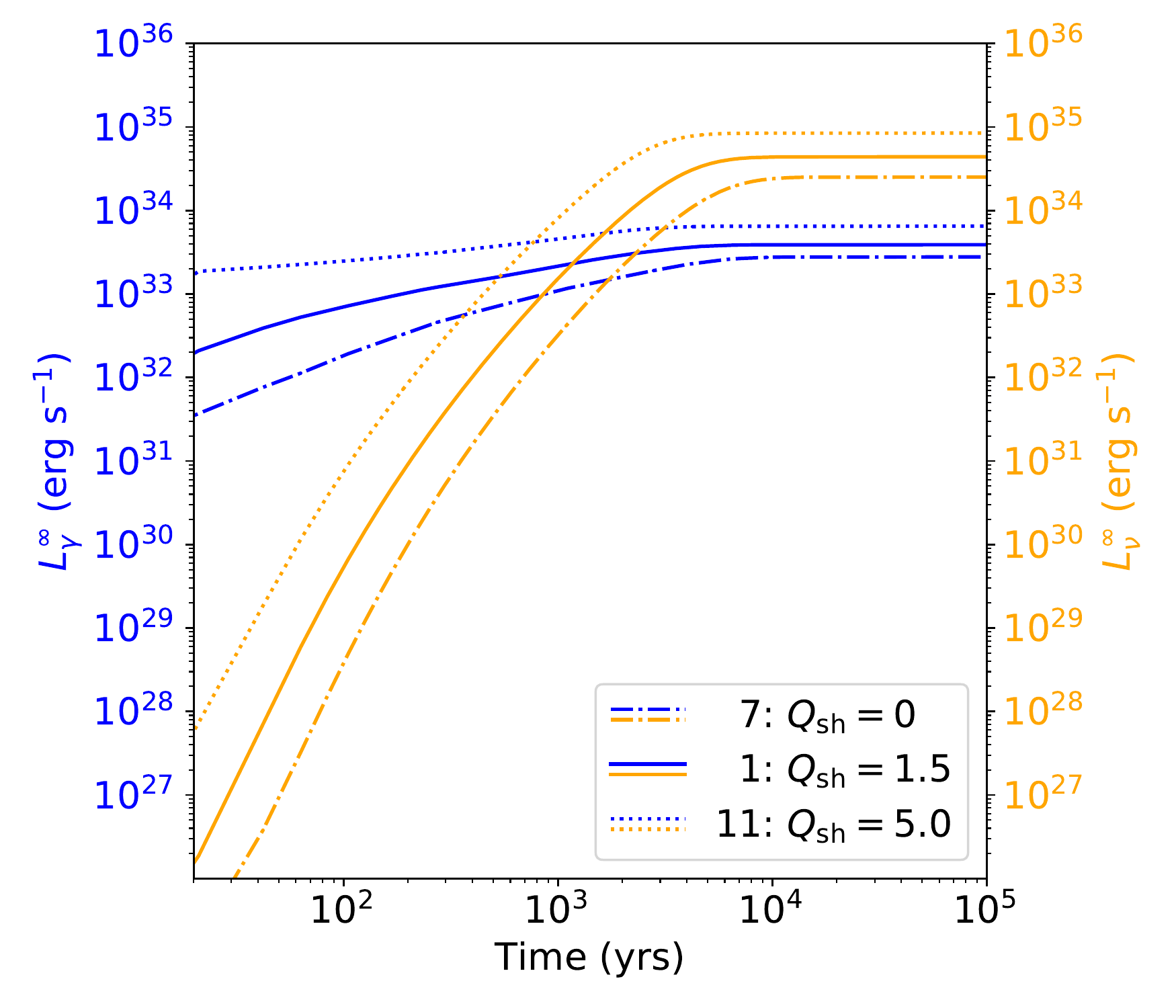}
     \caption{Same as Fig.~\ref{fig:trec}, but for models with different shallow heating strengths. In model 1 (solid curves) $Q_\text{sh}=1.5\text{ MeV nucleon}^{-1}$ is assumed. In model 7 (dashed-dotted curves) it is assumed that no shallow heating takes place and in model 11 (dotted curves) a shallow heating strength of $Q_\text{sh}=5.0\text{ MeV nucleon}^{-1}$ is taken into account. In all models, deep crustal heating is assumed to take place as well.}
     \label{fig:Qsh}
\end{figure*}

 \begin{figure*}
  \includegraphics[width=0.5\textwidth]{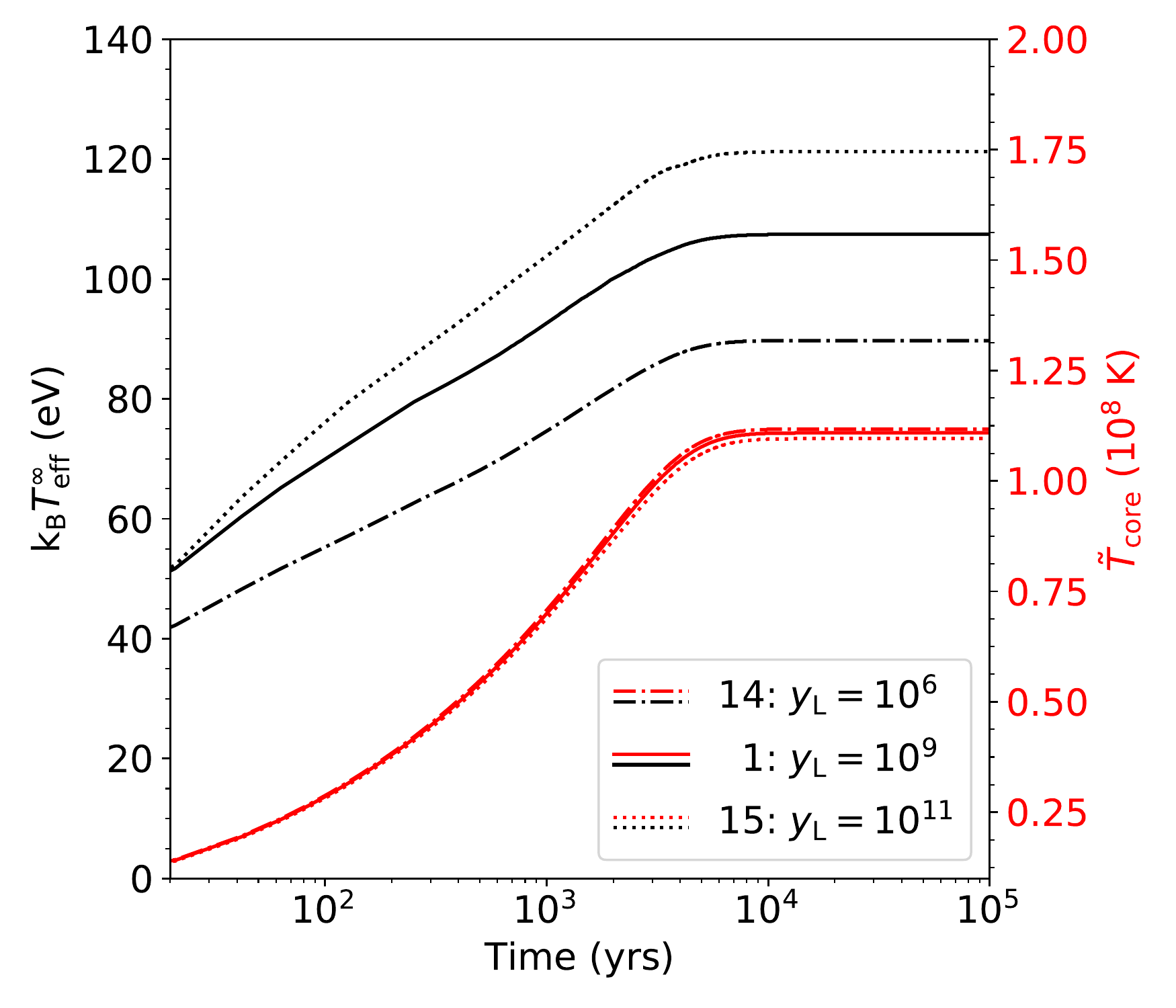}\includegraphics[width=0.5\textwidth]{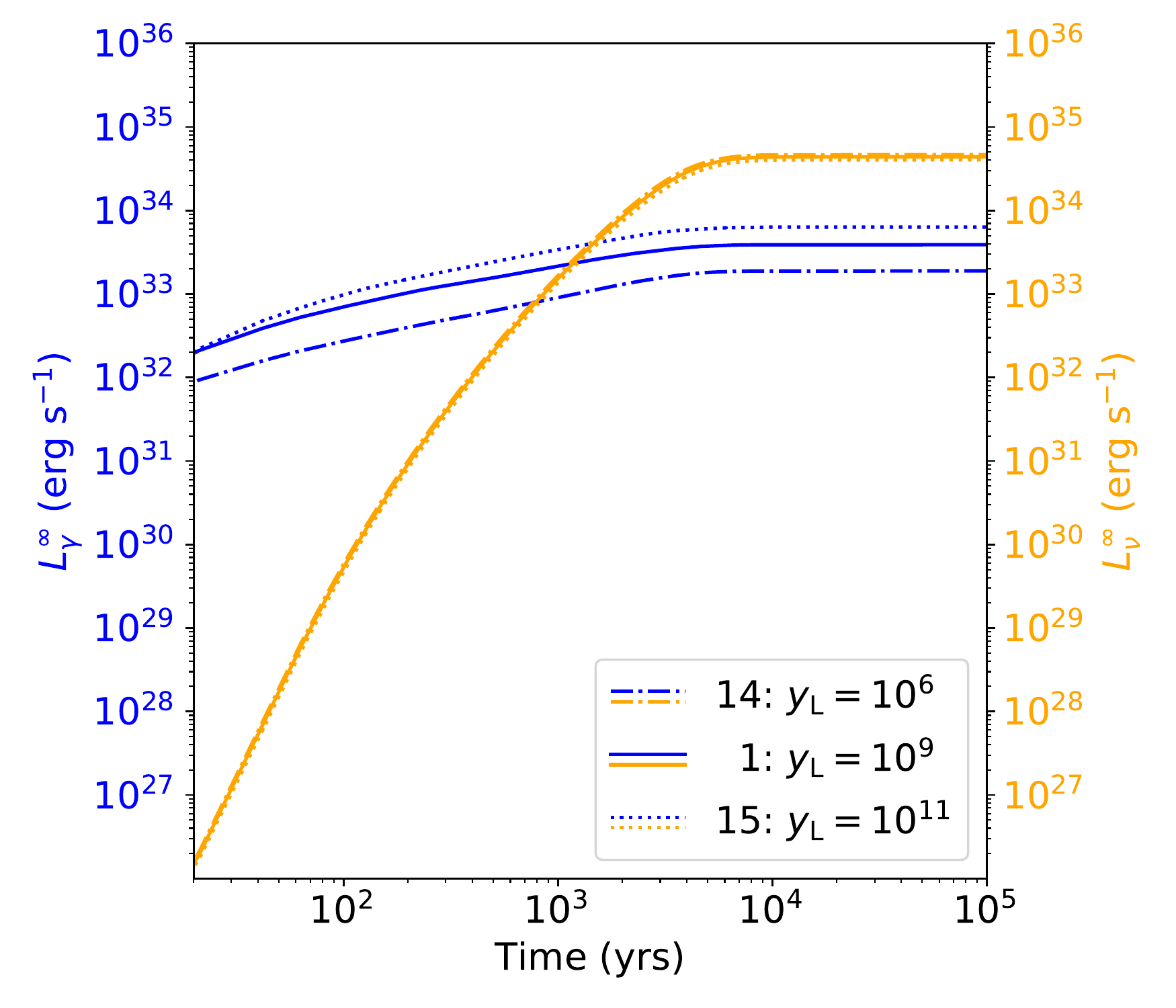}
     \caption{Same as Fig.~\ref{fig:trec}, but for different envelope compositions. In model 1 (solid lines), $y_\text{L}=10^9\text{ g cm}^{-2}$ is assumed, in model 14 (dash-dotted lines) we assume a heavy element envelope with $y_\text{L}=10^6\text{ g cm}^{-2}$, and in model 15 (dotted lines) we assume an envelope with a large light element column depth ($y_\text{L}=10^{11}\text{ g cm}^{-2}$). For a thicker light element column depth, the thermal conductivity in the envelope is higher, increasing the effective temperature ($T_\text{eff}(T_\text{b})$). At low temperatures ($\tilde{T}\sim10^7$~K) the $T_\text{b}-T_\text{eff}$ relation used here for models 1 and 14 converges.}
     \label{fig:env}
\end{figure*}

 \begin{figure*}
  \includegraphics[width=0.5\textwidth]{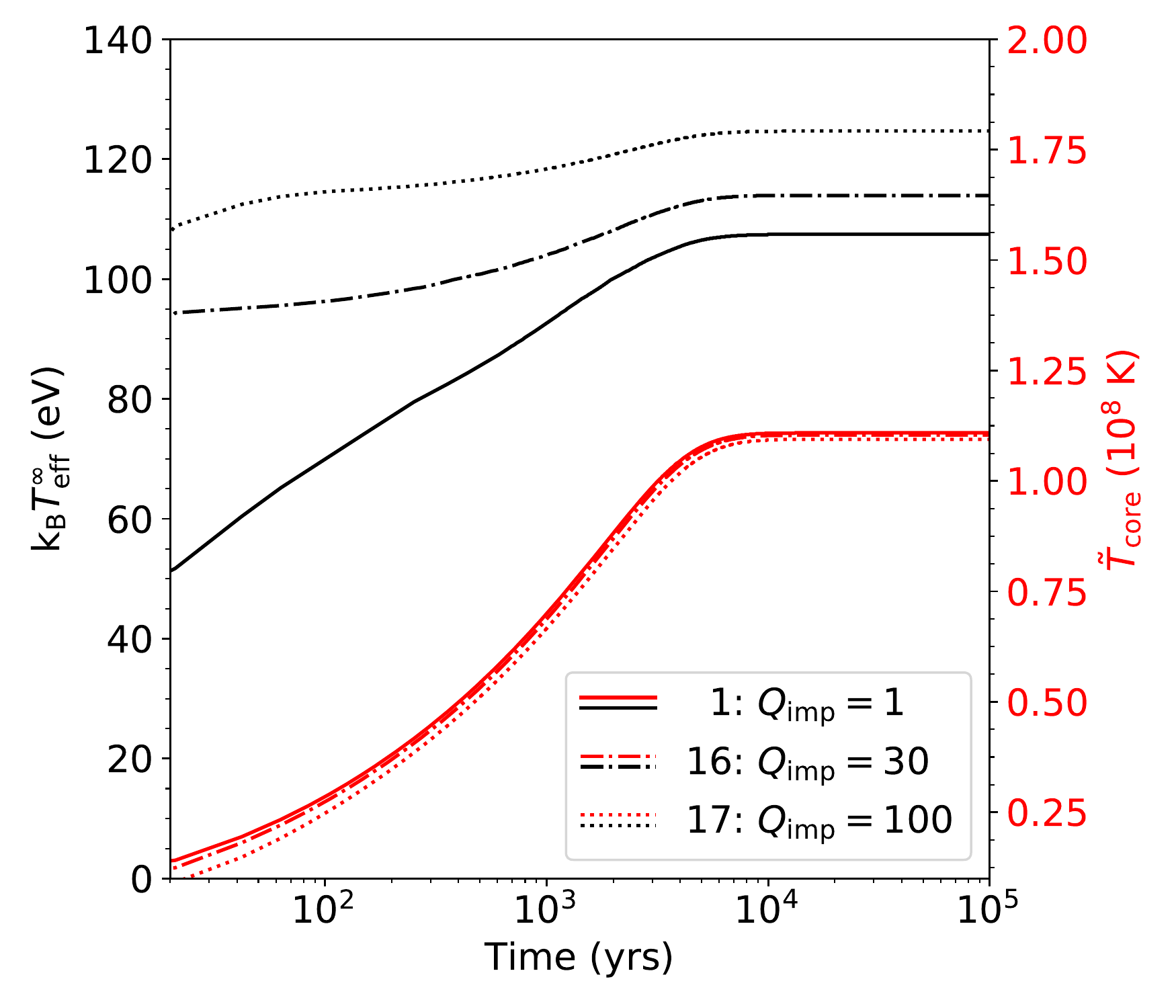}\includegraphics[width=0.5\textwidth]{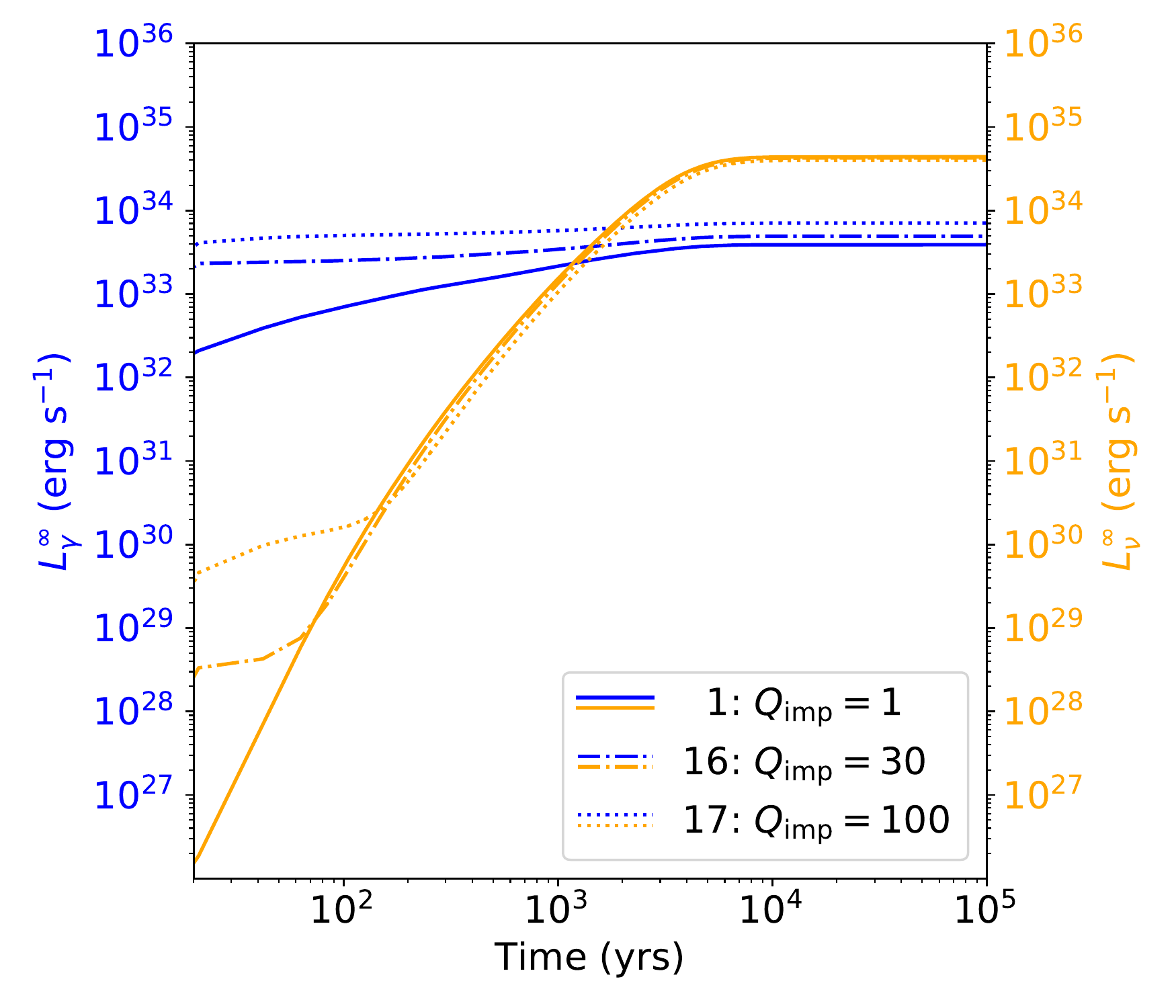}
     \caption{Same as Fig.~\ref{fig:trec}, but for different impurity factor in the (full) crust. In model 1 a crust with high thermal conductivity is assumed ($Q_\text{imp}=1$, solid lines), in model 16 the thermal conductivity in the crust is decreased by increasing the impurity parameter to 30 (dash-dotted lines), and $Q_\text{imp}=100$ is assumed throughout the crust in model 17. The plotted quantities here are taken when the neutron star is still in the crust-cooling phase, which causes the effective temperatures and photon luminosities for the models with lower thermal conductivity (16 and 17) to be higher, even though the core temperature in these models is lower. This is because for lower thermal conductivity, the crust takes longer to cool down after an outburst. If the quantities had been taken when the source was in crust-core equilibrium ($\sim10^{3-4}$ days after the end of an outburst), the neutron stars with lower core temperature would have lower surface temperatures and photon luminosities. }
     \label{fig:Qimp}
\end{figure*}

 \begin{figure*}
  \includegraphics[width=0.5\textwidth]{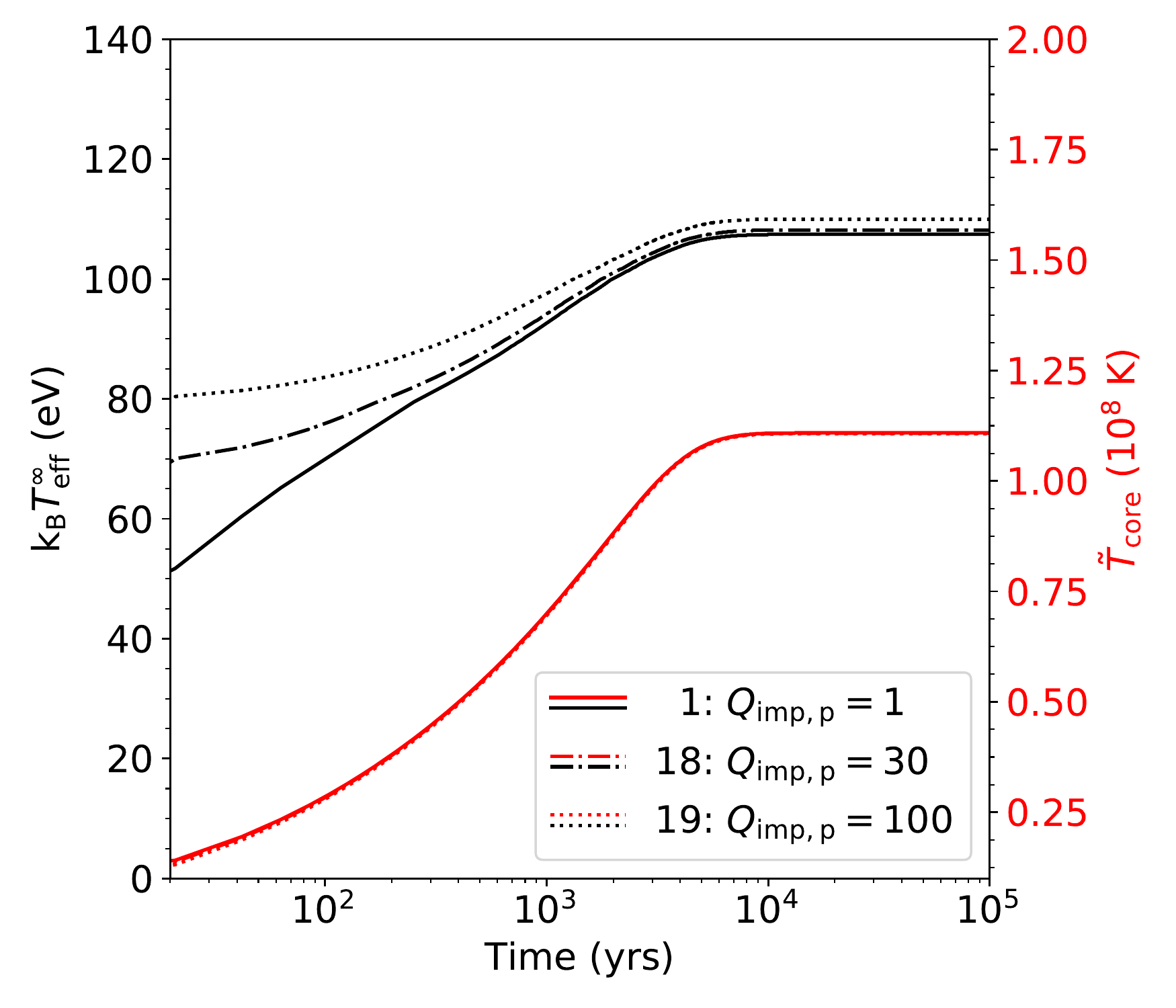}\includegraphics[width=0.5\textwidth]{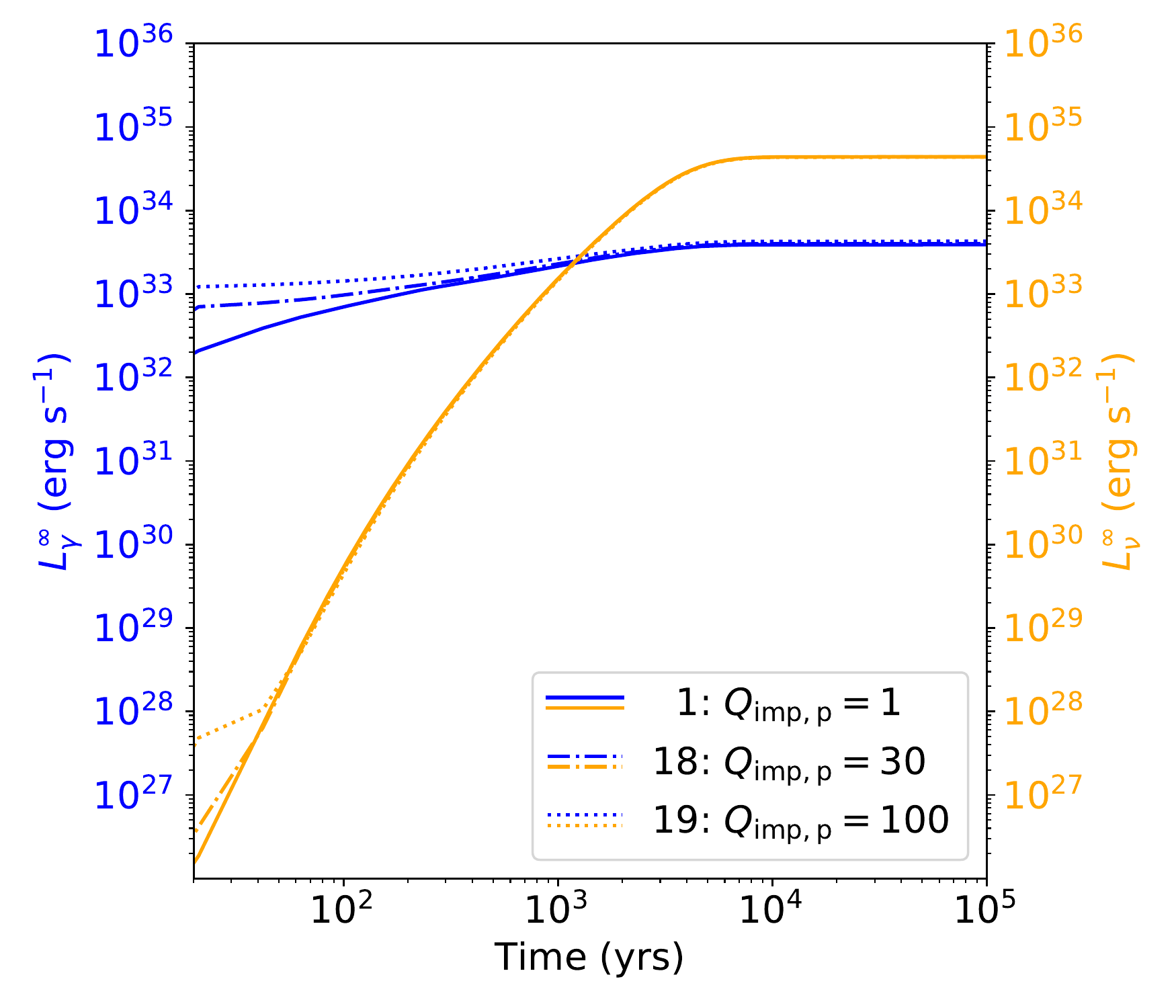}
     \caption{Same as Fig.~\ref{fig:Qimp}, but for different thermal conductivity in the pasta region ($\rho>8\times10^{13}\text{ g cm}^{-3}$). Outside the pasta region, the impurity factor in the crust is $Q_\text{imp}=1$ for all shown models. In model 1 an impurity factor of $Q_\text{imp,p}=1$ (solid curves) is assumed, in model 18 the impurity factor in the pasta region is set to $Q_\text{imp,p}=30$ (dash-dotted curves), and $Q_\text{imp,p}=100$ (dotted curves) is assumed in model 19. A higher impurity factor decreases the thermal conductivity (see Fig. \ref{fig:lambda}).}
     \label{fig:Qimppasta}
\end{figure*}

 \begin{figure*}
  \includegraphics[width=0.5\textwidth]{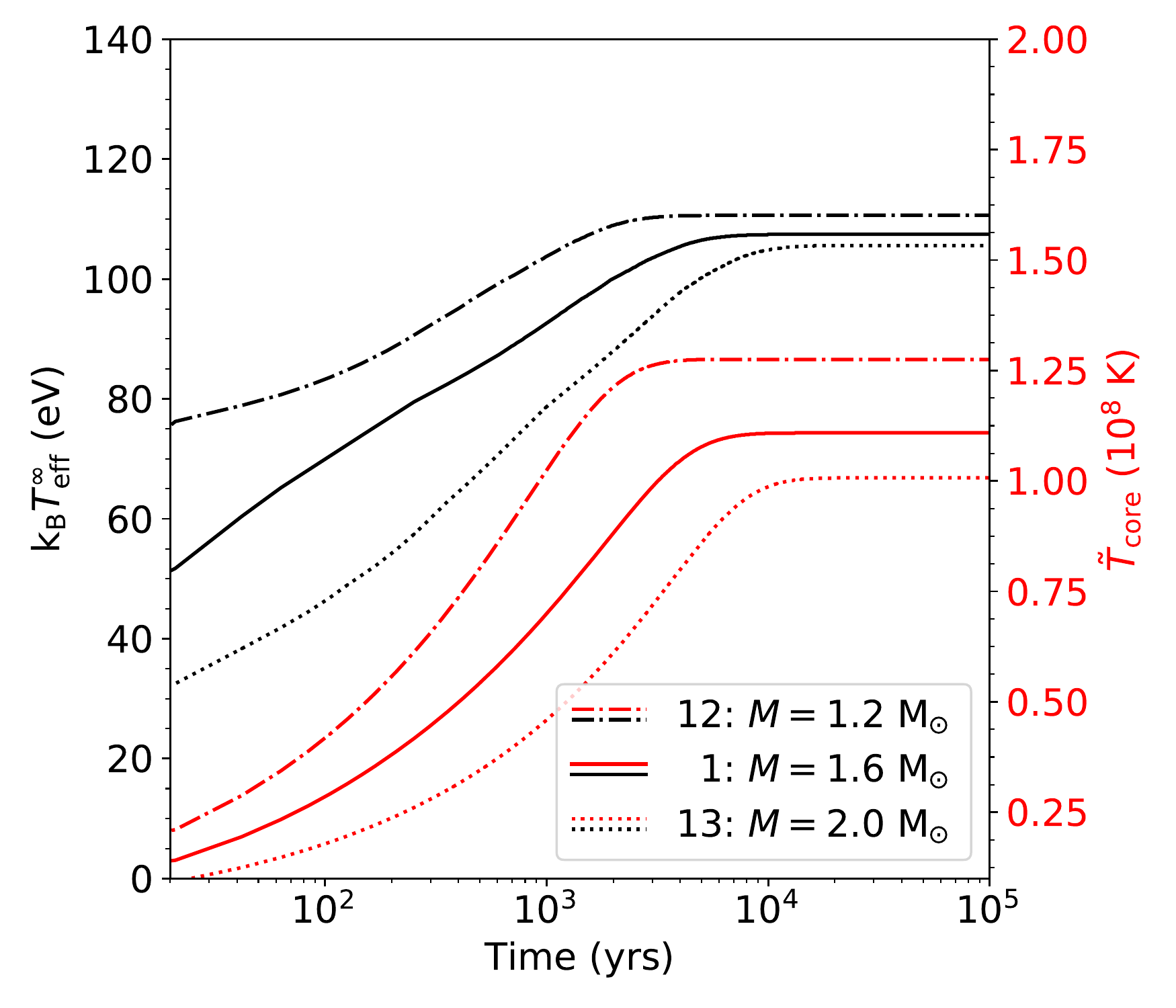}\includegraphics[width=0.5\textwidth]{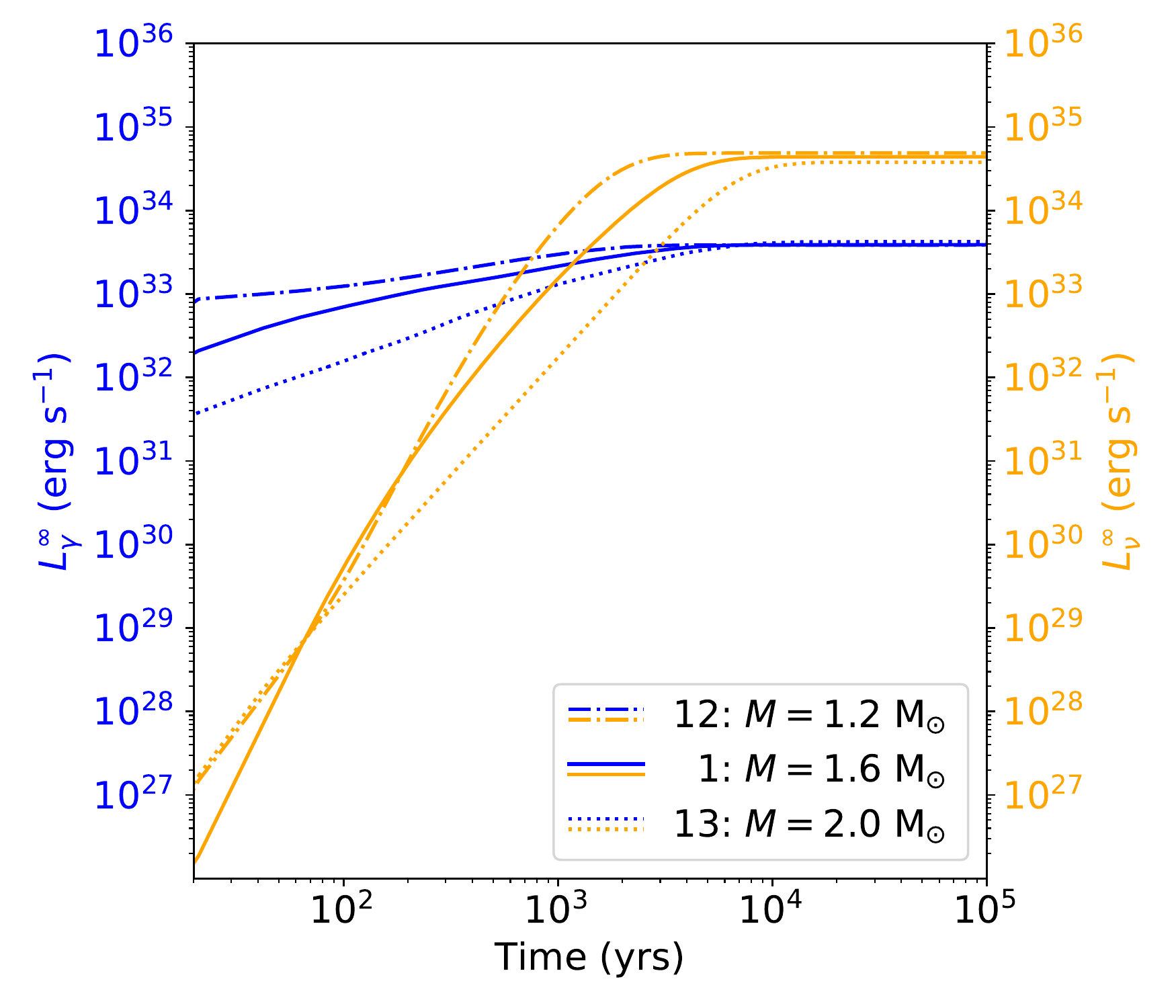}
     \caption{Same as Fig.~\ref{fig:trec}, but for models assuming different stellar masses (and consequently different radius and compactness, see Table~\ref{tab:masses}). In model 1 (solid curves) we assume ${M=1.6\text{ M}_\odot}$, a mass of ${M=1.2\text{ M}_\odot}$ was assumed in model 12 (dashed-dotted curves), and in model 13 we assume ${M=2.0\text{ M}_\odot}$ (dotted curves). In all models shown here we assume the cooling processes of the minimal cooling paradigm. Due to the changed mass, the neutron stars in these models also have different redshift factors (see Table~\ref{tab:masses}), which is why even though model 13 has the lowest surface temperature, it has the highest photon luminosity when the source has reached its long-term equilibrium state.}
     \label{fig:masses}
\end{figure*}

 \begin{figure*}
  \includegraphics[width=0.5\textwidth]{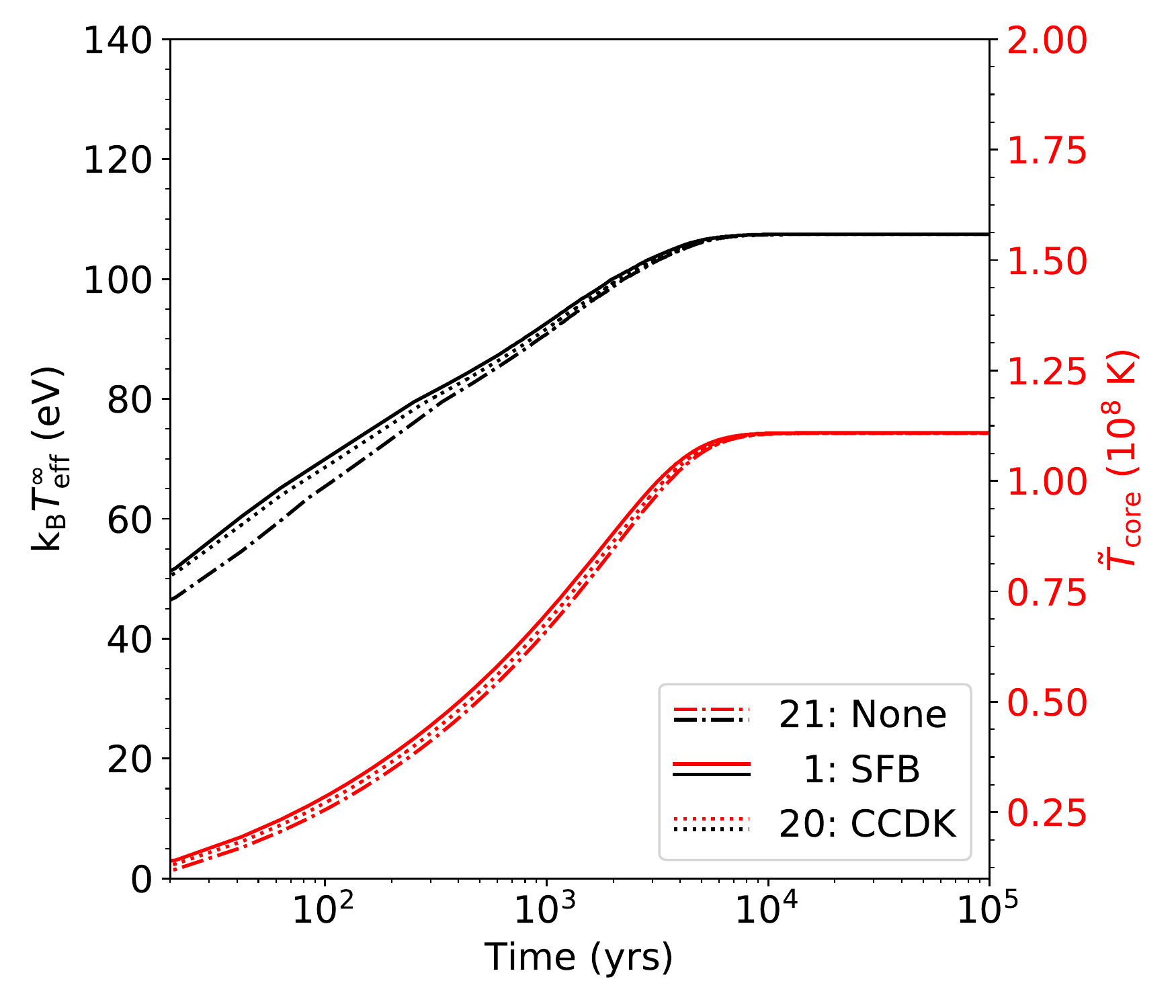}\includegraphics[width=0.5\textwidth]{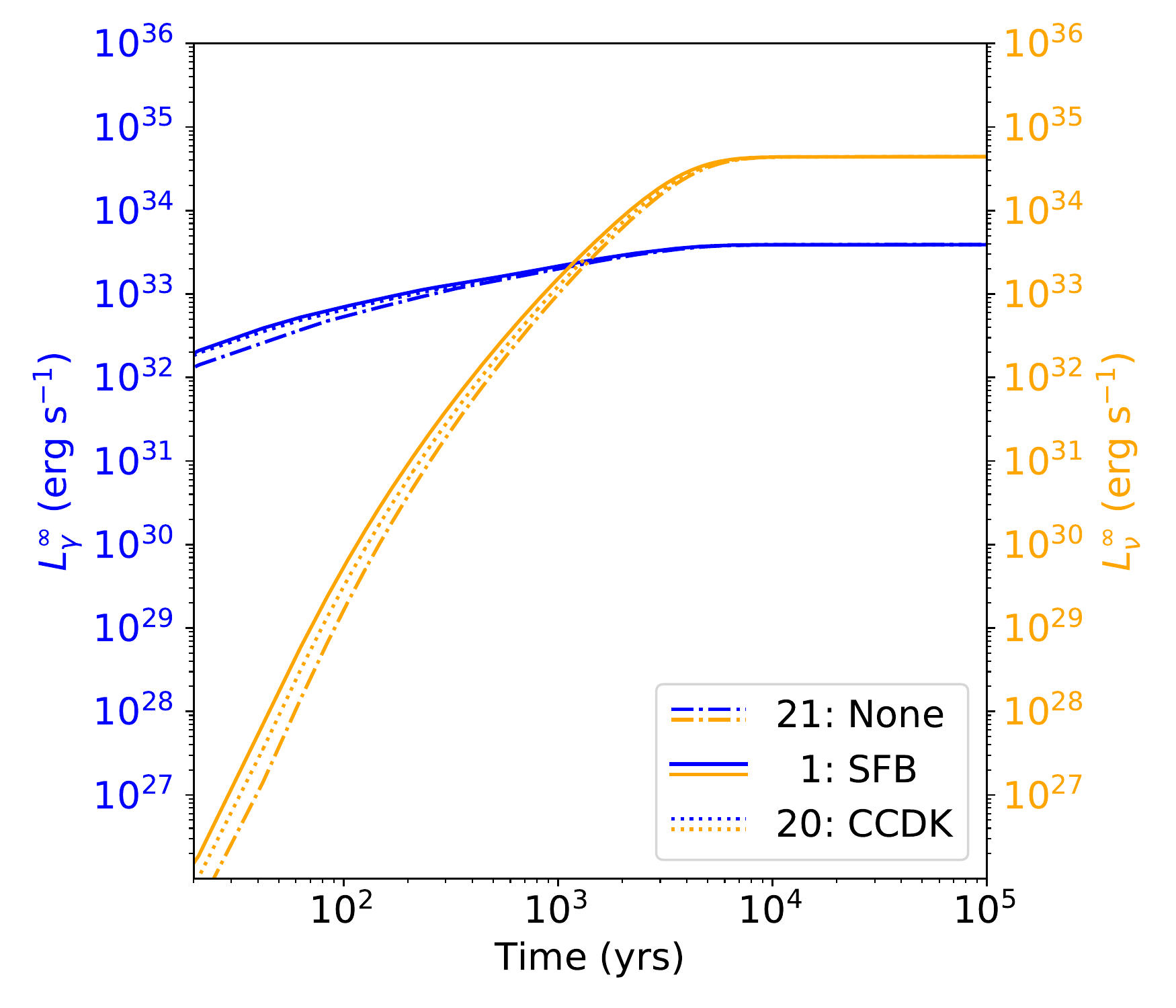}
     \caption{Same as Fig.~\ref{fig:trec}, but for models assuming different neutron $^1\text{S}_0$ pairing models. For curves labelled `SFB',  the neutron $^1\text{S}_0$ pairing gap from \citet{schwenk2003} is assumed (model 1). The label `CCDK' refers to the neutron $^1\text{S}_0$ gap model from \citet{chen1993} (model 20) and for the curves labelled `none' it is assumed that no neutron pairing takes place in the crust (model 21). All models assume the same neutron $^3\text{P}_2$ pairing gap.}
     \label{fig:n1s0}
\end{figure*}

 \begin{figure*}
  \includegraphics[width=0.5\textwidth]{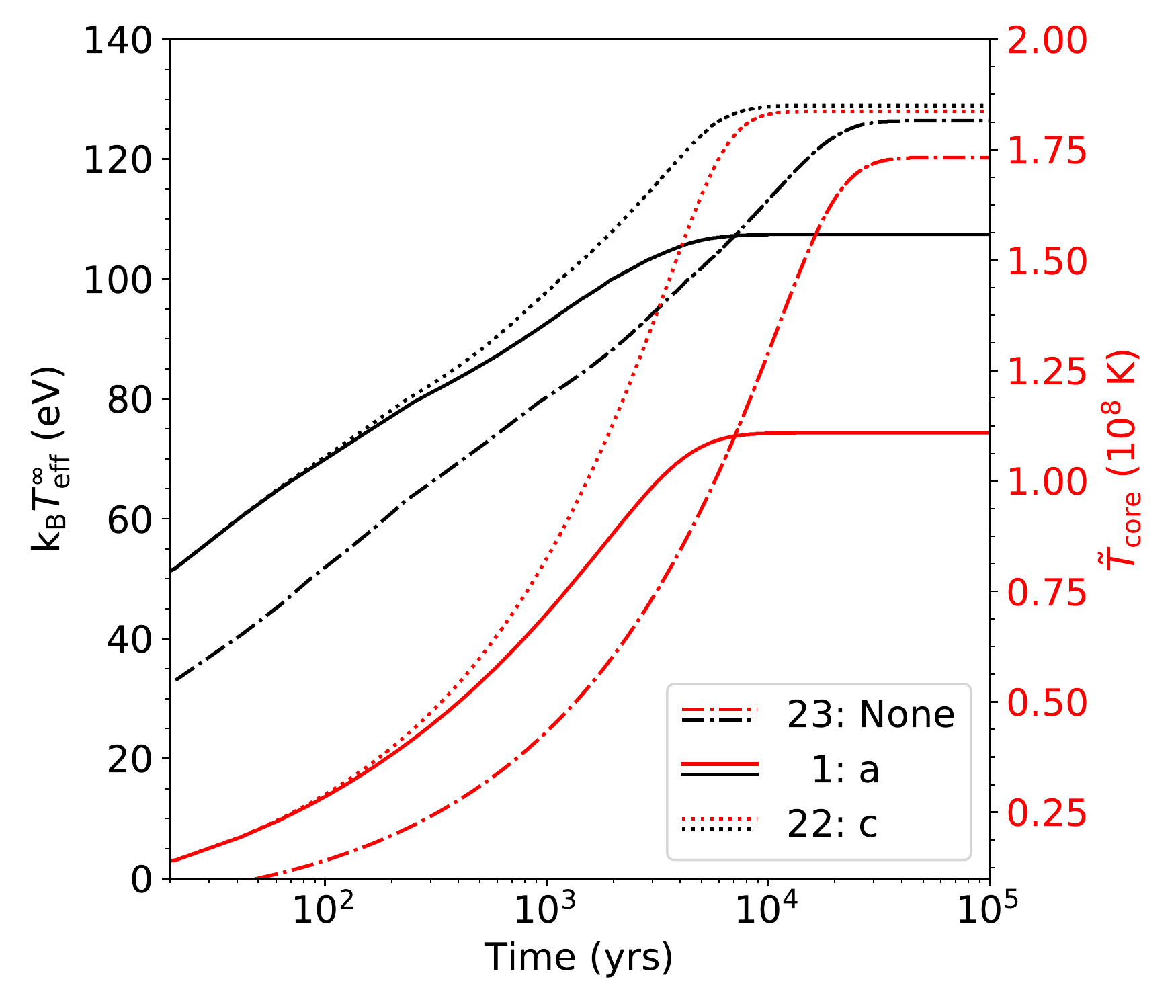}\includegraphics[width=0.5\textwidth]{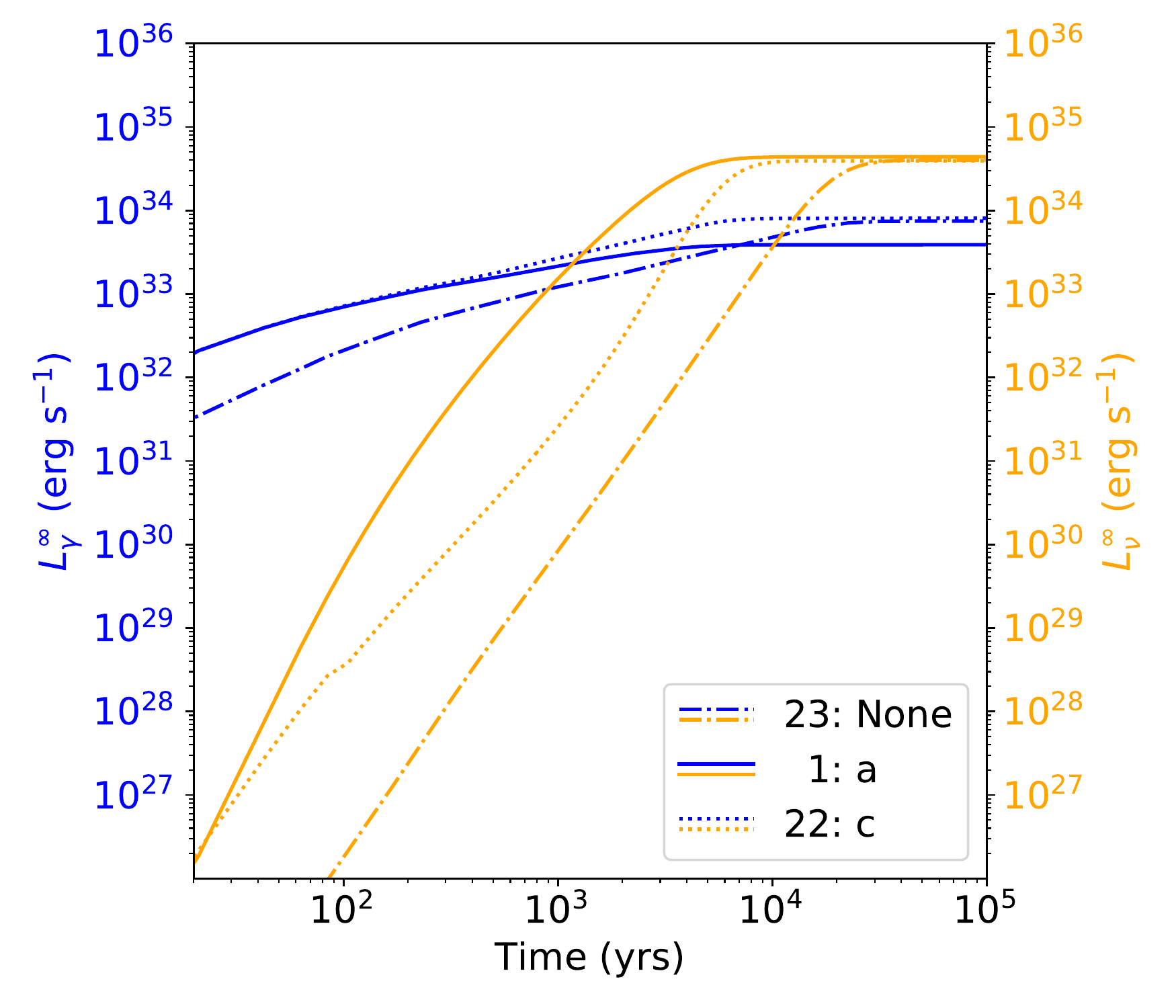}
     \caption{Same as Fig.~\ref{fig:trec}, but for models in which different neutron $^3\text{P}_2$ pairing gaps are assumed. Assumed here are neutron $^3\text{P}_2$ pairing gaps `a' (model 1, moderate pairing) and `c' (model 22, strong pairing) from \citet{page2004}, and we present the results of model 23 in which no neutron $^3\text{P}_2$ pairing is assumed to take place (curves labelled `none'). All models assumed the same neutron $^1\text{S}_0$ pairing gap. Figure \ref{fig:Lnu_n3p2} shows the neutrino luminosity as a function of temperature for these three models. }
     \label{fig:n3p2}
\end{figure*}

 \begin{figure*}
  \includegraphics[width=0.5\textwidth]{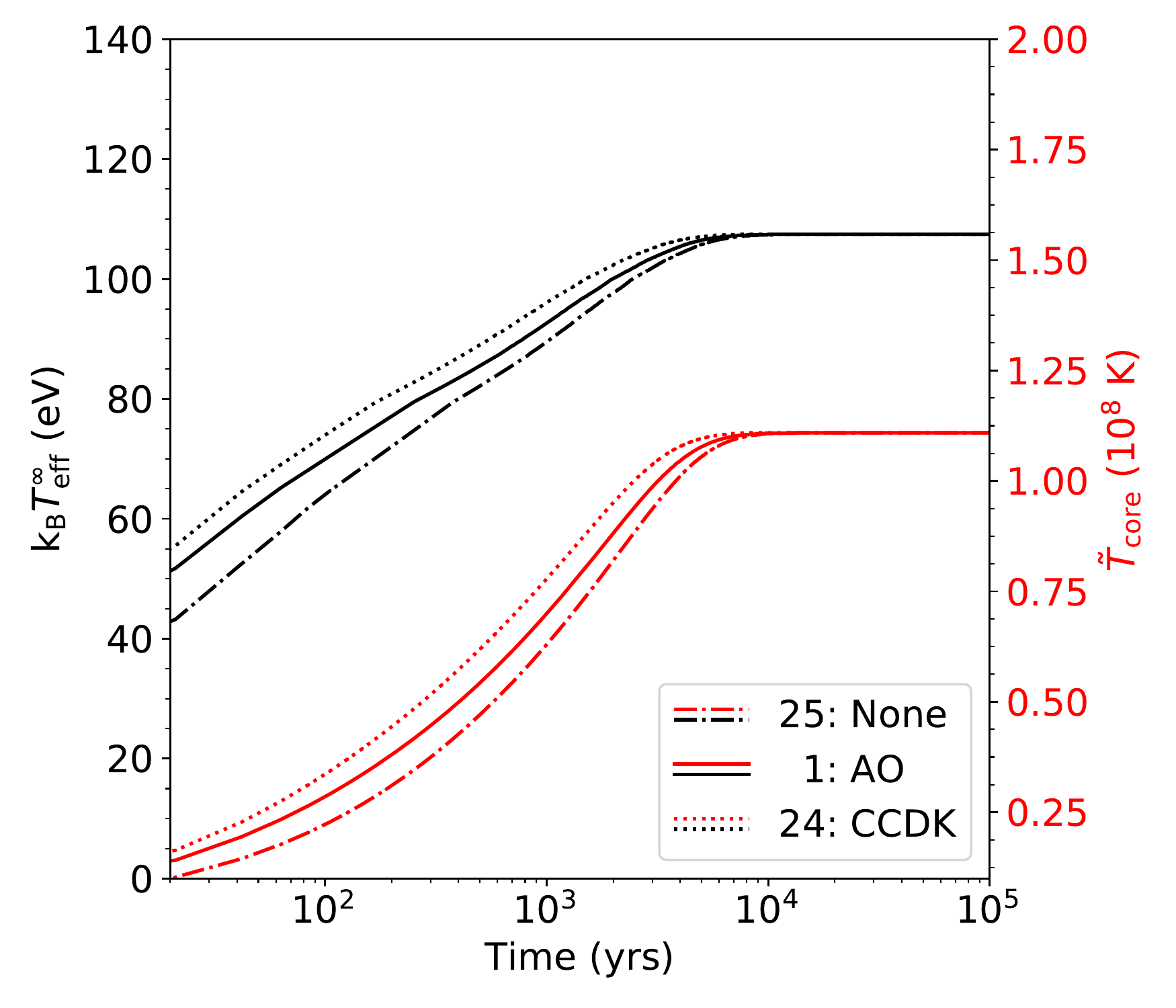}\includegraphics[width=0.5\textwidth]{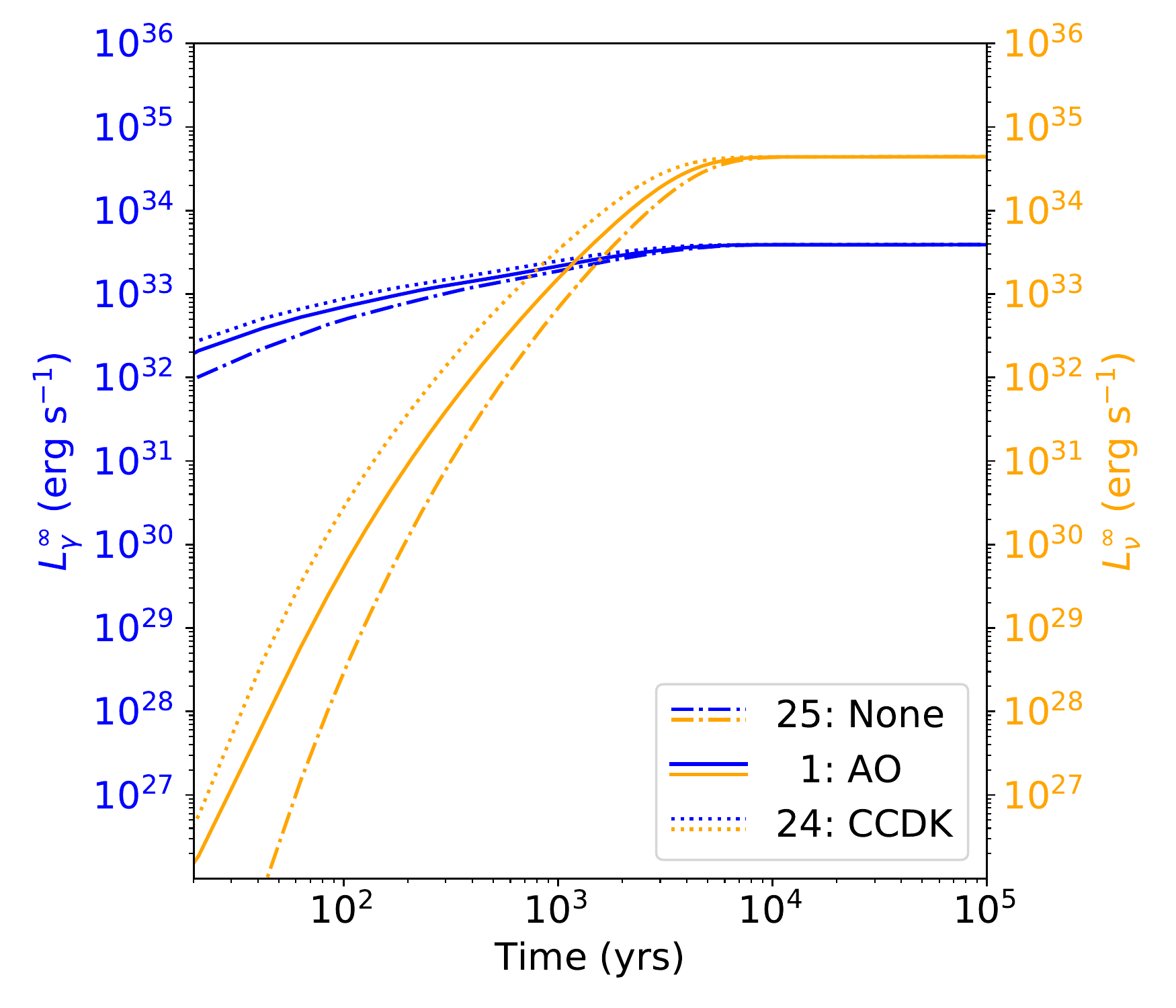}
     \caption{Same as Fig.~\ref{fig:trec}, but for models in which different proton $^1\text{S}_0$ pairing gaps are assumed. `AO' refers to the proton pairing gap from \citet{amundsen1985} as assumed in model 1. `CCDK' refers to the proton $^1\text{S}_0$ gap from \citet{chen1993}, assumed in model 24, and in model 25 (label `none') it is assumed that no proton pairing takes place. }
     \label{fig:p1s0}
\end{figure*}

 \begin{figure*}
  \includegraphics[width=0.5\textwidth]{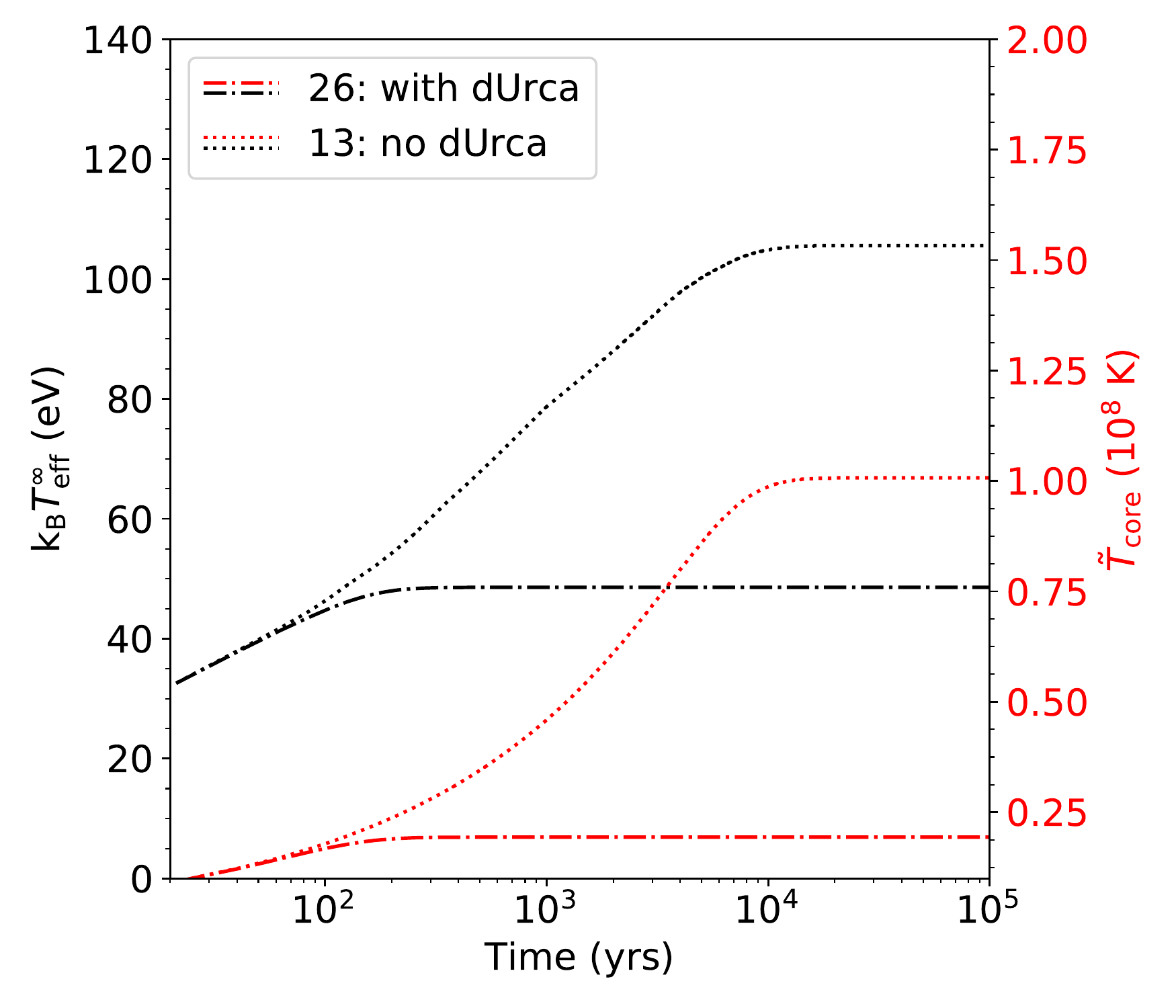}\includegraphics[width=0.5\textwidth]{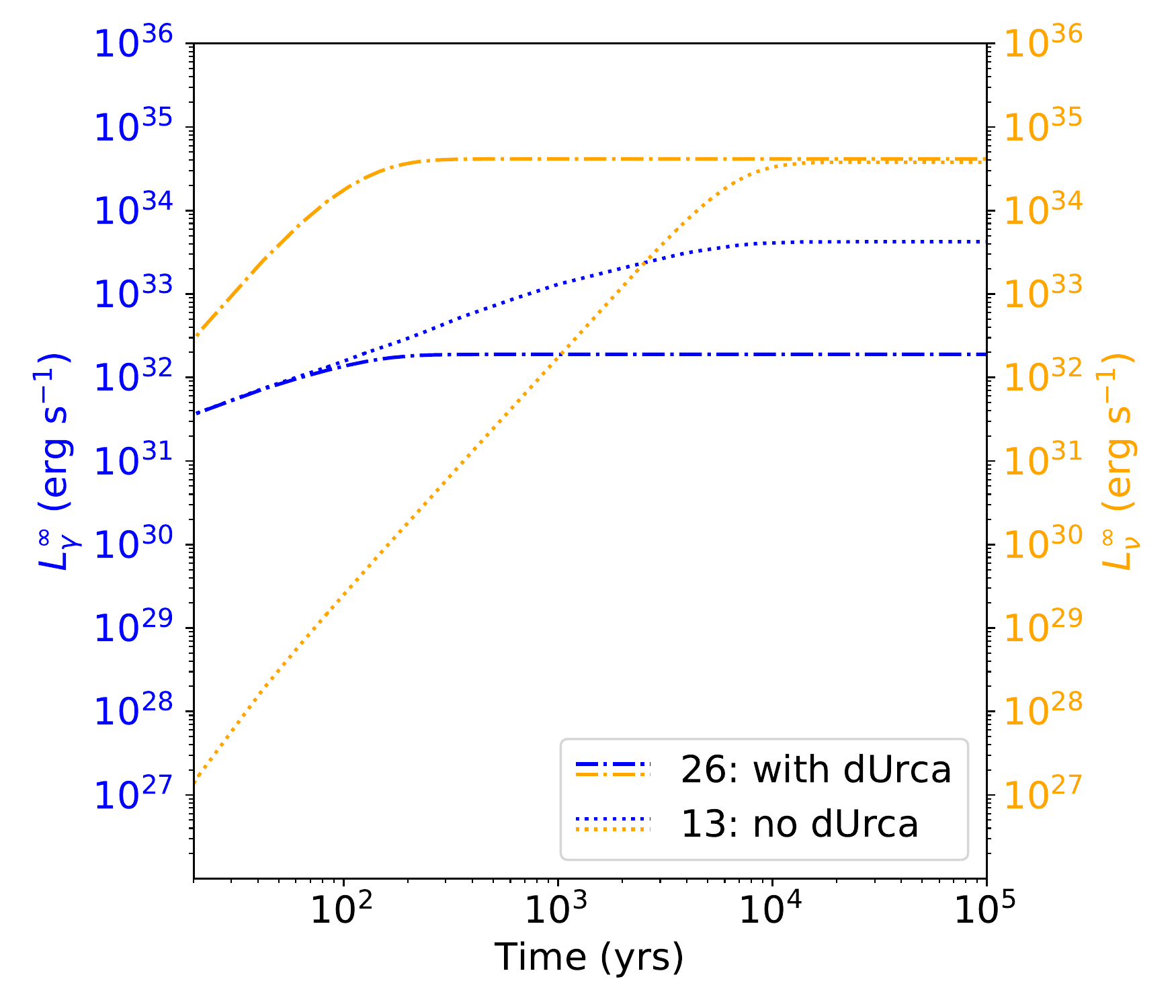}
     \caption{Same as Fig.~\ref{fig:trec}, but for two models in which a neutron star with mass $M=2.0\text{ M}_\odot$ is assumed. In model 26 the direct Urca process takes place at densities $\rho\geq1.257\times10^{15}\text{ g cm}^{-3}$. To show the effect that this neutrino emission process has on the temperature evolution of the star, the direct Urca process is prohibited to take place in model 13 (label `no dUrca'). Figure~\ref{fig:lumdurca} shows the neutrino luminosities as a function of temperature as used in {\tt NSCool} for model 26.}
     \label{fig:durca}
\end{figure*}

\end{document}